\documentclass[a4paper,fleqn]{cas-sc}

\usepackage[numbers]{natbib}
\usepackage{lscape} 
\usepackage{graphicx}
\usepackage{overpic}
\usepackage{subfigure}
\usepackage{dcolumn}
\usepackage{multirow}
\usepackage{hyperref}
\hypersetup{
    colorlinks = true,
    urlcolor   = blue,
    linkcolor = blue,
    citecolor  = blue,
}
\def\tsc#1{\csdef{#1}{\textsc{\lowercase{#1}}\xspace}}
\tsc{WGM}
\tsc{QE}

\begin{document}
\let\WriteBookmarks\relax
\def\floatpagepagefraction{1}
\def\textpagefraction{.001}

\shorttitle{LES of hydrodynamic noise from turbulent flows past an axisymmetric hull using high-order schemes}    

\shortauthors{P. Jiang et al.}  

\title [mode = title]{Large-eddy simulation of hydrodynamic noise from turbulent flows past an axisymmetric hull using high-order schemes}

\author[1]{Peng Jiang}[orcid=0000-0002-9072-8307]
\credit{Methodology, Software, Validation, Writing – original draft, Formal analysis, Visualization}

\author[1,2]{Shijun Liao}[orcid=0000-0002-2372-9502]
\credit{Conceptualization, Supervision, Project administration, Funding acquisition}

\author[1,2]{Bin Xie}[orcid=0000-0002-4218-2442]
\credit{Conceptualization, Supervision, Writing – review and editing, Funding acquisition}

\cortext[1]{Corresponding author}
\ead{xie.b.aa@sjtu.edu.cn}

\affiliation[1]{organization={School of Ocean and Civil Engineering, Shanghai Jiao Tong University},
                city={Shanghai},
                postcode={200240}, 
                country={China}}

\affiliation[2]{organization={State Key Laboratory of Ocean Engineering, Shanghai Jiao Tong University},
                city={Shanghai},
                postcode={200240}, 
                country={China}}

\begin{abstract}
In this paper, wall-modeled large-eddy simulation (WMLES) is carried out with Ffowcs-Williams and Hawkings (FW--H) acoustic analogy to investigate the turbulent flow and hydrodynamic noise of an axisymmetric body of revolution. We first develop the numerical model based on high-order schemes and validate it by benchmark test of the turbulent flow around a circular cylinder at \(\mathrm{Re}=1.0 \times 10^4\). It demonstrates the capability of the present scheme to capture the primary flow patterns and the acoustic noise in the far field.
Then, we conduct the numerical simulation for the turbulent flows around the DARPA SUBOFF without appendages at the Reynolds number of \(\mathrm{Re} = 1.2 \times 10^7\). The numerical results such as pressure coefficients and velocity fluctuations, are accurately predicted by the present model, which shows closer agreement with the experimental data than available WMLES solutions in the literature. 
\textcolor{black}{For the parallel midbody of the hull, the wall pressure fluctuation reveals a low-frequency broadband spectrum with the majority of signal energy. The surface fluctuating pressure spectrum scales to the power of Strouhal number at the different locations, which is consistent with those in the turbulent boundary layer of the plate flows and airfoils.} \textcolor{black}{Moreover, the acoustic signature in the far field is investigated where the lowest sound pressure level (\textit{SPL}) occurs in the upstream and downstream directions while the highest is found in the mid-parallel plane. \textit{SPLs} are relatively close in the region of high acoustic pressure at the transverse plane of \(x/D=0\), which exhibits a maximum difference of 1.2 dB between locations at different angles. In the vertical plane at \(z/D=0\), the directivity plot reveals a symmetrical dipole pattern with vertical fluctuations stronger than the streamwise fluctuations.}

\end{abstract}


\begin{keywords}
large-eddy simulation \sep turbulent noise \sep high-order schemes \sep axisymmetric hull \sep acoustic analogy
\end{keywords}

\maketitle
\section{Introduction}\label{introduction}

Flow-induced noises are common in nature. The noises generated by water flow around ships and other underwater vehicles are serious concerns for various naval engineering applications. Additionally, turbulence-generated noise can detrimentally affect the stealth operations of a vehicle, potentially masking the faint signals targeted by passive sonar systems \citep{Wang2006noiseReview,Yu2007shipNoise,Li2016review,Zhou2022suboff, He2024Review}. However, accurate prediction of turbulent flow and noise generated by submarines remains challenging using numerical approaches for several reasons: (i) The flow around a submarine model involves intricate phenomena, including boundary layer transition, flow separation, turbulent boundary layers, adverse pressure gradient zones, and wake flow; (ii) Complex geometries and curved boundaries pose challenges in generating high-quality meshes for numerical simulations; (iii) High Reynolds number (\(\mathrm{Re} \sim 10^{7-8}\)) introduces spatial and temporal multiscale complexities, making the near-wall flow scale difficult to resolve and demanding significant computational resources; (iv) Low Mach number (\(\mathrm{Ma} \sim 0.01\)) results in a substantial difference between the characteristic scales of the flow and acoustic fields, leading to relatively small noise energy \citep{Wang2006noiseReview}. Against this background, this paper attempts to address this challenging issue. This work focuses on the turbulent noise of flow past an axisymmetric hull at a high Reynolds number. The hull of the DARPA SUBOFF model (AFF1) \citep{groves1989geometric} is considered to gain insights into the hydrodynamic noise produced by the turbulent flows around more intricate underwater vehicles in engineering applications such as submarine hulls and torpedoes.
    
Predicting flow-induced noise is a challenging task that needs an accurate and efficient approach to characterize and understand the sound source in the region of turbulence. In recent years, several numerical and theoretical techniques have been established to determine the flow-induced noise of flows passing over fixed or moving bodies. These include semi-empirical, direct, and hybrid approaches \citep{Wang2006noiseReview,Robison2014propeller,Boukharfane2020WallPressure}. Acoustic analogy is a commonly used computational method for predicting flow-induced noise within the framework of hybrid methods. 
The foundational work for the acoustic analogy was laid by \citet{Lighthill1952I, Lighthill1954II}, who reformulated governing equations to derive a wave equation for sound propagation. A critical aspect in numerically predicting flow-induced noise lies in identifying an effective sound source within the flow. In Lighthill's acoustic theory, various sound sources are taken into consideration, including the dynamic coupling of fluctuating velocity, entropy variations, and viscous stress.
\citet{Curle1955} further developed a formulation to solve the Lighthill acoustic equation by considering the contribution of solid surface to the sound field. The contributions are concluded as two parts which are (i) the reflection and diffraction of the acoustic waves at the solid surface and (ii) the generation of a resulting dipole field at the solid surface.
Additionally, \citet{FWH1969} expanded Curle's solution to incorporate arbitrarily moving bodies, resulting in the FW--H equation. The FW--H analogy proves applicable for acoustic sources in relative motion to a rigid surface, a scenario frequently encountered in various applications such as naval engineering. The computation typically includes quadrupole, dipole, and monopole terms. Building on this, \citet{Farassat20071A} then developed solutions to the FW--H equation specifically for subsonic moving surfaces, denoted Formulations 1 and 1A, which are different in the treatment of time derivatives.
Currently, studies of the hydro-acoustics of underwater vehicles, such as ship propellers, submarine hulls, and hydrofoils, usually rely on the acoustic analogy within the framework of hybrid methods due to its appealing balance between numerical accuracy and computational efficiency. This methodology is developed based on the assumption that the hydrodynamics are negligibly affected by the acoustic results, allowing the hydrodynamics to be resolved without consideration of the influence from hydroacoustics \citep{Posa2023propeller}. Subsequently, acoustic pressure can be reconstructed from the obtained fluid dynamics results. 

The solution fidelity of the hybrid approach hinges strongly on the turbulence model used for the numerical simulation. In predicting the time-averaged fluid dynamics and forces of submarine propellers \citep{Chase2013OE} and the self-propulsion performance of submarines \citep{Sezen2018OE}, the Reynolds-averaged Navier-Stokes (RANS) methodology has proven successful. Nevertheless, it poses challenges in predicting noise because of insufficient information on the fluctuating flow field since the RANS method only yields averaged velocities and pressures, with flow fluctuations modeled using Reynolds stress models.
Alternatively, large-eddy simulation (LES) directly resolves the dynamics of the energy-dominant vortex and uses subgrid-scale (SGS) models to describe the influence of the turbulent scales which can not be resolved by the mesh \citep{Boukharfane2020WallPressure}. It has been shown to outperform the RANS model as the small scales typically exhibit greater isotropy and homogeneity than the larger scales, making them more appropriate for universal modeling.
Depending on whether the turbulent scales are resolved or modeled in the near-wall region, LES can be further divided into wall-resolved large-eddy simulation (WRLES) and wall-modeled large-eddy simulation (WMLES). The computational cost of WRLES is much more demanding for the inner layer since quasi-parallel coherent structures in the flow direction have a strong influence on the near-wall flow dynamics, especially for high-Reynolds number cases. \citet{Choi2012gridWMLES} demonstrated that the total grids required to resolve the inner part of the boundary layer (\(y^+ = y u_\tau /\nu < 100\), and \(y\) is the direction normal to the wall) scales to \(\mathrm{Re}^{13/7}\). 
For example, \citet{posa2016suboff} investigated the fully appended SUBOFF model at \(\mathrm{Re} = 1.2 \times 10^6\) using WRLES with 208 million grid nodes and reported a shift in the peak of turbulent kinetic energy from the hull, resulting in a double-peaked wake stress shape. Subsequently, \citet{Kumar2018SUBOFF} conducted WRLES on the SUBOFF hull at \(\mathrm{Re} = 1.1 \times 10^6\) employing 608 million control volumes, observing similar turbulence behavior to the mid-parallel body of the hull compared to flat-plate turbulent boundary layer (TBL). Later, \citet{morse2021suboff} conducted WRLES and streamlined coordinate analysis of the SUBOFF hull at \(\mathrm{Re} = 1.1 \times 10^6\) using 712 million control volumes. To the best of the authors' knowledge, most previous studies on the hydrodynamics of submarines are primarily confined to \(Re=1.2 \times 10^6\) using WRLES due to the limitation of computational resources. In a more recent study, \citet{LIU2023112009} utilized 1.476 billion grid cells to fully resolve the near-wall flows for large-eddy simulations of flows around an underwater vehicle model with $\mathrm{Re_L} = 1.2 \times 10^6$. An exception is the work by \citet{Qu2021suboff}, which employed a relatively coarser grid (approximately 108 million grid nodes) to study the large-scale vortex of flow past the SUBOFF model at \(Re=1.2 \times 10^7\) by LES without wall models. However, for more realistic engineering applications, turbulent flows around underwater vehicles, such as submarine hulls and torpedoes, and particularly in the field of hydroacoustics, involve higher Reynolds numbers (\(\mathrm{Re} \sim 10^{7-8}\)). The size of the near wall vortices relative to the thickness of the boundary layer decreases with increasing Reynolds number. This introduces resolution requirements to resolve the spatial and temporal multiscale of turbulence, which may exceed current computational capabilities.

A more cost-effective alternative is available from the WMLES approach, which is becoming increasingly important in engineering research. In this method, the wall boundary conditions are augmented to account for the effect of the unresolved inner layer of the boundary layer while the majority of the boundary layer is resolved by LES. For example, \citet{Boukharfane2020WallPressure} utilized equilibrium wall-modeled LES, achieving commendable agreement with experimental data for surface pressure fluctuations and trailing-edge noise of an airfoil. They concluded that equilibrium WMLES provides a potential approach to investigate the turbulent pressure field induced by the TBL for accurate and efficient simulation. \citet{posa2020suboff} conducted equilibrium WMLES for flow past the appended SUBOFF model at \(\mathrm{Re} = 1.2 \times 10^7\), demonstrating that an increase in Reynolds number induces a small peak in turbulence energy in the outer layers above the tail. Simultaneously, the circulation of the transition vortex and its contribution to turbulence intensify at higher Reynolds numbers. \textcolor{black}{More recently, in the work of \citet{Zhou2022suboff}, a second-order scheme with an immersed boundary method (IBM) is employed to explore the turbulent noise of flow past the SUBOFF model with appendages at \(\mathrm{Re} = 1.2 \times 10^7\) using non-equilibrium WMLES and acoustic analogy. The main focus is on the directivity of the far-field noise in four transverse planes around the SUBOFF. Their study revealed that WMLES serves as a possible approach for predicting flow characteristics and far-field noise directivity of submarine vehicles.} \citet{Rocca2022BB2} employed a second-order finite volume method (FVM) model implemented in OpenFOAM to study the flow past BB2 submarine model at \(\mathrm{Re} = 1.2 \times 10^6\). The study utilized advective FW--H equations with WMLES, with a specific focus on the acoustic near-field. Their results showed that the massive flow separations in the wake region are the main source of the near-field noise, and the sound pressure level exhibits a low-frequency broadband spectrum. \textcolor{black}{Additionally, \citet{Chen2023SUBOFF} and \citet{He2023SUBOFF} performed WMLES with a second-order FVM method for turbulent flows around the SUBOFF bare hull at \(\mathrm{Re} = 1.2 \times 10^7\), demonstrating that the non-equilibrium wall stress model can effectively predict flow characteristics for the wall-bounded flow problems.}

All of these applications suggest the potential of the WMLES methodology for predicting high-Reynolds number turbulent flows and far-field acoustics. However, it is worth noting that in nearly all of the studies reported above, the
numerical method was limited to second-order schemes. It is unclear whether using WMLES with high-order methods in complex engineering problems is more advantageous than traditional low-order models, particularly for studying turbulence-generated noise. Moreover, the accuracy of far-field acoustic fields depends on both the fidelity of fluid dynamics simulations and the conservation properties of the numerical method employed for resolving the governing equations \citep{Posa2023propeller}. Additionally, the simulation of the turbulent flows is accompanied by difficulties related to structures across multiple scales in space and time. It is reported by \citet{xie2017piso} that traditional second-order schemes have excessive numerical dissipation which could mask the result of the SGS model in large-eddy simulation. Consequently, sufficient small mesh size and time steps are required to guarantee convergence and accuracy, which prohibitively increases the computational cost. To enhance the numerical accuracy as well as efficiency of simulation, \citet{Xie2019High-fidelitysolver,Xie2020consistent} proposed the finite volume method based on a merged stencil with 3rd-order reconstruction (FVMS3) on the polyhedral unstructured grids. 
Several numerical tests have been carried out to demonstrate that this scheme offers third-order accuracy and substantially reduces undesired numerical dissipation and dispersion and thus resolves more elaborate flow structures.
While high-order models have proven advantageous in direct numerical simulation (DNS) of low and medium Reynolds number turbulence problems, the effectiveness of combining high-order schemes with WMLES for solving complex high-Reynolds-number engineering turbulence problems remains unclear. To tackle this issue, it is essential to investigate the potential integration of the turbulence model and high-order scheme via the numerical simulation of some basic problems. Despite extensive research on numerical simulations of high Reynolds flow around submarine hulls in recent years, the analysis of hydroacoustics and sound directivity of submarine hulls based on high-order simulations using a WMLES approach has been largely unexplored, with the exception of work by \citet{Zhou2022suboff,Rocca2022BB2}. Consequently, this study aims to fill this gap and address the following questions: (i) Can high-order schemes with WMLES provide absolute advantages for solving complex high-Reynolds-number turbulence problems? (ii) What are the characteristics of wall pressure fluctuations for flow past the SUBOFF hull? (iii) What is the far-field acoustic pressure distribution in all spatial directions for the SUBOFF hull?

To address these questions, we first develop a hybrid numerical model that combines the WMLES and FW--H equations using the high-order FVMS3 scheme. Then, we perform the numerical simulation of turbulent noise around the cylinder at a subcritical Reynolds number, which are compared with existing experimental and numerical results. After the validation, this model is used to investigate the turbulent flow past the SUBOFF hull at Reynolds number of \(\mathrm{Re} = 1.2 \times 10^7\) which marks an initial step toward a more in-depth quantitative investigation of far-field noise of underwater vehicles based on high-order WMLES. Additionally, high-performance computing is utilized to reconstruct the acoustic far-field from the WMLES data, considering 648 hydrophones placed at a distance of 500 diameters from the SUBOFF hull. These results provide a comprehensive depiction of acoustic pressure distribution in all spatial directions, representing a wealth of information not present in previous literature.

Below, the paper is organized as follows: Section~\ref{methodology} gives a brief explanation of the methodology, including the solution of fluid dynamics, wall-stress modeling, FW--H acoustic analogy, and the hybrid method for acoustic simulation. Section~\ref{numerical_validation} presents numerical validation for flow past a cylinder at $\mathrm{Re}=1.0\times10^4$, including flow characteristics, flow validation, and validation with experimental far-field noise results. Section~\ref{suboff} discusses WMLES of turbulent flow and noise around an axisymmetric hull, which includes numerical configurations, details of acoustic post-processing, instantaneous vortex structure, flow characteristics, and acoustic analysis of numerical results. Finally, Section~\ref{conclusions} draws the conclusions of the paper.

\section{Methodology}\label{methodology}
\subsection{Governing equations}\label{governing_equations}
The LES approach is adopted in this work, for which the large-scale turbulence is resolved and a subgrid-scale model is used to compute the effect of unresolved scales of turbulence. The governing equations are the spatially filtered incompressible Navier--Stokes equations, given as
\begin{gather}
    \frac{\partial \widetilde{u}_i}{\partial x_i}=0, \label{eq-cont}
\\[3pt]
    \frac{\partial \widetilde{u}_i}{\partial t}+\frac{\partial \widetilde{u}_i \widetilde{u}_j}{\partial x_j}=-\frac{\partial \widetilde{p}}{\partial x_i}-\frac{\partial \tau_{i j}}{\partial x_j}+\nu \frac{\partial^2 \widetilde{u}_i}{\partial x_j \partial x_j},\label{eq-momnt1}
\end{gather}
where $i$ and $j$ are indices that span the three directions in space, the tilde symbol $\widetilde{\cdot}$ denotes the spatial filtering over the grid in Cartesian coordinates, $x_i$ represents the coordinate in direction $i$, $\widetilde{u}_i$ is the filtered velocity in space along the direction $i$, $\widetilde{p}$ is the filtered pressure divided by the density, and $\nu$ is the kinematic viscosity. In Eq.~\eqref{eq-momnt1}, the term $\tau_{ij} = \widetilde{u_i u_j}-\widetilde{u}_i \widetilde{u}_j$ is the SGS stress tensor. Additionally, the deviatoric part is given by
\begin{equation}
	\tau_{i j}^d=\tau_{i j}-\frac{\tau_{k k}}{3} \delta_{i j}=-2 v_t \widetilde{S}_{i j},
 \label{eddy-viscosity}
\end{equation}
where $\widetilde{S}_{i j}$ is the strain rate tensor of the resolved field, $\tau_{kk}$ is the trace of the SGS stress tensor, $\delta_{i j}$ the Kronecker delta and $\nu_t$ the eddy-viscosity. For incompressible flows, by substituting Eq.~\eqref{eddy-viscosity} into Eq.~\eqref{eq-momnt1}, the momentum equation can be rewritten as follows
\begin{equation}
	\frac{\partial \widetilde{u}_i}{\partial t}+\frac{\partial \widetilde{u}_i \widetilde{u}_j}{\partial x_j}=-\frac{\partial \widetilde{p}}{\partial x_i}+\nu_{\mathrm{eff}} \frac{\partial^2 \widetilde{u}_i}{\partial x_j \partial x_j},\label{eq-momnt2}
\end{equation}
here, the total viscosity, denoted as $\nu_{\mathrm{eff}}$, is the sum of the kinematic viscosity ($\nu$) and the eddy viscosity ($\nu_t$). \textcolor{black}{This study addresses the turbulence closure problem by employing the wall-adaptive local eddy-viscosity (WALE) model \citep{Nicoud1999WALE}. }

\subsection{Wall-stress modelling}\label{wall_model}
In WMLES, the outer part of the TBL is resolved by a coarser computation mesh. However, the coarser LES grid is not sufficient to capture the strong velocity gradients, the dynamics of the stress-carrying coherent structures, and the strong momentum exchange in the near-wall region using only the SGS models.
\textcolor{black}{Wall modeling methods aim to model the dynamics of the inner layer and correct for numerical inaccuracies in the underresolved TBL of LES by directly enhancing the wall stresses by using the modeled stress for the LES to the wall \citep{Hayat2023WMLES}.} Hereafter, we provide a brief description of the wall model employed in this study, coupled with the WMLES method.

\begin{figure}[htbp]
	\centering
	\begin{overpic}[width=15cm]{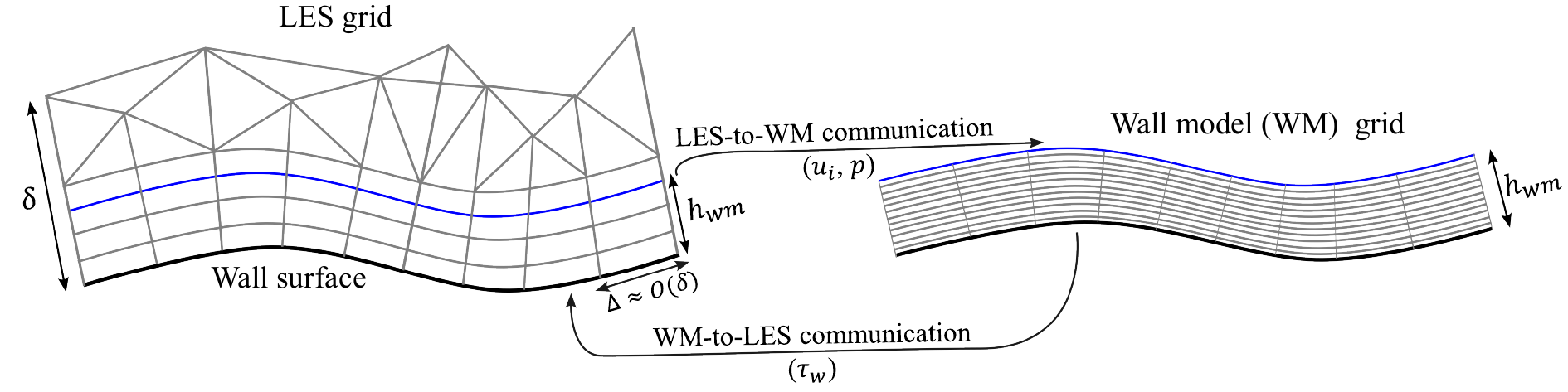}
	\end{overpic}		
	\caption[]{Schematic of wall model routine coupled with the LES solver.}
	\label{fig:WMLES_schematic}
\end{figure}

Figure~\ref{fig:WMLES_schematic} shows a diagram of the WMLES procedure. During each simulation step, there are two communications between the LES and the wall models (WM) to advance the simulation: the LES-to-WM communication and the WM-to-LES communication. To facilitate LES-to-WM communication, the LES calculates the required flow variables ($u_i,p$) for wall models stored in LES cells at the corresponding matching height ($h_{wm}$). These flow variables are then used as input for the calculation of wall stresses $(\tau_w)$ by the wall models. \textcolor{black}{In the WM-to-LES communication, the modeled stress calculated by the WM is used to enhance the boundary condition of the wall-adjacent cell. Within the OpenFOAM framework, it corrects the effective eddy viscosity $(\nu_{\mathrm{eff}})$ for the wall boundary condition using the wall shear stress $(\tau_w)$ from the WM.} 

Wall models work out the simplified turbulent boundary layer equations (TBLEs), also known as the Ordinary Differential Equation (ODE)-type wall model, on a separate finer grid.  The ODE-type wall model necessitates only wall-normal discretization, rendering it locally one-dimensional. This characteristic simplifies implementation and reduces computational costs. Specifically, this study employs an ODE-type wall model that takes into account non-equilibrium effects arising from pressure gradients:

\begin{gather}
    \frac{\partial}{\partial x_2}\left[\left(\nu+\nu_t\right) \frac{\partial\left\langle u_i\right\rangle}{\partial x_2}\right]=F_i,
    \label{eq:TBLE}
\\[3pt]
    F_i=\frac{1}{\rho} \frac{\partial\langle p\rangle}{\partial x_i},
\end{gather}
where the index \(i\) takes values \(1\) and \(3\) representing the two wall-parallel directions and $\langle \cdot \rangle$ denotes the time averaging. The wall-normal component (\(i = 2\)) can be determined from the continuity equation. Consequently, the equations need to be solved within the region \(x_2 = 0\) at the wall to \(x_2 = h_{wm}\) at the sampling location from the LES. The formulation presented in Eq.~(\ref{eq:TBLE}) has incorporated the Boussinesq assumption regarding the Reynolds stress tensor. Integrating Eq.~(\ref{eq:TBLE}) analytically yields the solution for wall shear stress:
\begin{equation}
    \left\langle\tau_{w, i}\right\rangle=\left(\left.\left\langle u_i\right\rangle\right|_ {h_{wm}}-F_i \int_0^{h_{wm}} \frac{x_2}{\nu+\nu_t} \mathrm{~d} x_2\right) / \int_0^{h_{wm}} \frac{\mathrm{d} x_2}{v+v_t}.
    \label{eq:wall_shear_stress}
\end{equation}
The wall shear stress \(\left\langle\tau_{w}\right\rangle\) in Eq.~(\ref{eq:wall_shear_stress}) can be determined based on the provided \(F_i\) and \(\langle u_i \rangle\) at the sampled location (\(h_{wm}\)) from the simulation. The eddy-viscosity \(\nu_t\) is computed by the numerical model developed by \citet{Duprat2011nutmodel}, which has demonstrated superior performance in the numerical simulation of turbulent flows with separation. 

Subsequently, the computed value of the wall shear stress, denoted by $\left\langle\tau_{w, i}\right\rangle$, is enforced at the face center in order to update the effective wall eddy viscosity, which is represented by $\nu_{\mathrm{eff}}$:
\begin{equation}
   \langle \tau_{w,i} \rangle = \nu_{\mathrm{eff}} \frac{\left\langle u_{i,p}\right\rangle}{\Delta x_2}, 
    \label{eq:UpdateNuEff}
\end{equation}
where the index \(i\) takes values \(1\) and \(3\) representing the two wall-parallel directions. The subscript `$p$' indicates that the evaluation is to be carried out at the center of the wall adjacent to the face center of the wall. The wall-normal distance between the two aforementioned points is designated by the symbol $\Delta x_2$. In the present work, the wall shear stress is predicted using the dynamic library developed by \citet{Mukha2019WMLES}.


\subsection{Ffowcs-Williams \& Hawkings acoustic analogy}\label{FWH_analogy}
\textcolor{black}{
The most formal solution to Lighthill’s analogy is the FW--H formulation \citep{FWH1969}, which considers the influence of solid boundaries with arbitrary motion. The integral form of the FW--H equation typically includes the surface and volume integrals, which are also known as quadrupole, dipole, and monopole terms. \citet{Wang2006noiseReview} and \citet{Brentner2003rotors} have given a thorough review of the mathematical foundations for the FW--H formulation.  Since the focus of this study is on the acoustic far field, the volume integrals can be neglected \citep{Zhou2022suboff, Posa2023propeller}. Thus, this study reconstructs the acoustic far field by computing only the dipole or loading component, which is the sum of the surface integrals of the FW--H equation. Furthermore, this study focuses specifically on problems involving stationary solid surfaces as observed by the observer. Thus, the surface integral term associated with the motion of the solid body can also be neglected. For simplicity, we present here the necessary terms of FW--H for the present study:
\begin{equation}
4 \pi \hat{p}(\boldsymbol{x}, t)=  \frac{1}{c} \frac{\partial}{\partial t} \int_{\mathbb{S}}\left[\frac{p^{\prime} \widehat{n}_i \widehat{r}_i}{r\left|1-\mathbb{M}_r\right|}\right]_{\mathbb{T}} d S  +\int_{\mathbb{S}}\left[\frac{p^{\prime} \widehat{n}_i \widehat{r}_i}{r^2\left|1-\mathbb{M}_r\right|}\right]_{\mathbb{T}} d S.
\label{eq:FWH_equation}
\end{equation}
In Eq.~(\ref{eq:FWH_equation}), on the left-hand side (LHS), \(\hat{p}\) denotes the sound pressure. \(\boldsymbol{x}\) denotes the position vector of the observer of the sound waves, and \(t\) represents time. On the right-hand side (RHS), the surface at position vector \(\boldsymbol{y}\) in the source region are represented, respectively, by $\mathbb{S}$. \(c\) is the speed of sound in the particular fluid, \(\boldsymbol{r} = \boldsymbol{x}-\boldsymbol{y}\) denotes the distance between the source and the receiver, and the scalar \(r\) represents the magnitude of the vector \(\boldsymbol{r}\). 
Additionally, \(\widehat{r_i}\) represents the component in the direction \(i\) in space of the unit vector \(\boldsymbol{r}/|\boldsymbol{r}|\), and similarly, \(\widehat{r_j}\) denotes the component in the direction \(j\), and $\widehat{n_i}$ is the component in the direction \(i\) in space of the unit vector \(\widehat{\boldsymbol{n}} \), normal to the component dS of the surface $\mathbb{S}$. $\mathbb{M}_r$  denotes the flow Mach number in the $\mathbf{r}$ direction. Note that the Ma is small for the present work on hydroacoustics.
Here, \(p^*\) and \(\rho^*\) denote the reference pressure and reference density which are equal to the free-stream pressure and fluid density respectively in this case. 
All terms on the RHS of Eq.~(\ref{eq:FWH_equation}) should be calculated at the emission time $\mathbb{T}$. However, for hydroacoustics, the time delay is typically negligible $\mathbb{T}\approx t$, as demonstrated in earlier works on the subject \citep{Cianferra2019shipNoise, Posa2022propeller}. Therefore, this study ignores the emission time, as considering the time delay would unnecessarily complicate the acoustic post-processing. For the present problem, the surface of integration, $\mathbb{S}$, is chosen on the bodies' surfaces.}

In the subsequent discussion of the acoustic signature, the \textit{SPLs} will be expressed in decibels, defined as:
\begin{equation}
\textit{SPL}=20 \log _{10}\left(A_{\text{FFT}}[\widehat{p}] / \widehat{p}_0\right),
\label{eq:SPL}
\end{equation}
where \(A_{\text{FFT}}[\widehat{p}]\) represents the amplitude of the acoustic pressure after Fast Fourier Transform (FFT) at a specific frequency. The term \(\widehat{p}_0\) denotes a typical reference pressure, set to \(1 \, \mu\text{Pa}\) for underwater radiated noise which is defined by the hydroacoustics of flow over an axisymmetric hull.

\subsection{Solution procedure of hybrid method}\label{High-order_schemes} 

\begin{figure}[htbp]
	\centering
	\begin{overpic}[width=16cm]{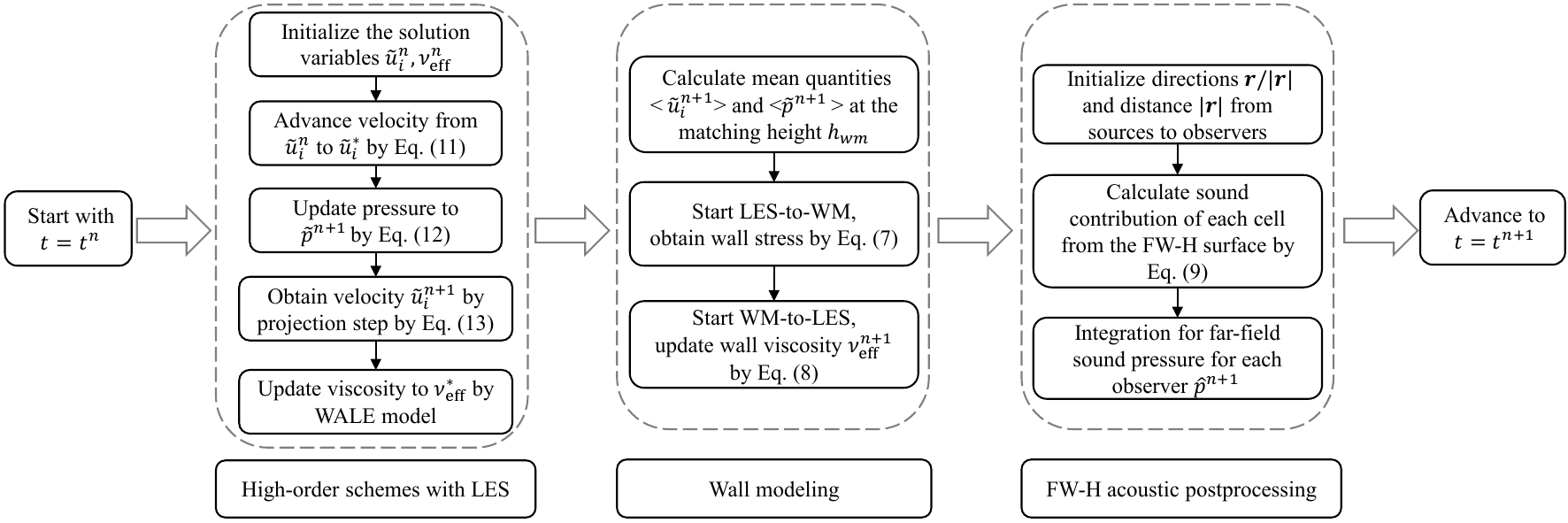}
	\end{overpic}		
	\caption[]{A flow chart for the whole solution procedure of the present hybrid method in a time step.}
	\label{fig:flow_chart_hybrid_method}
\end{figure}

To investigate turbulent flow and associated far-field turbulent noise for flow around an axisymmetric hull in a computationally efficient manner, a hybrid simulation approach is employed. This hybrid simulation consists of two parts: the WMLES framework to resolve turbulent boundary layers and wakes near the hull and the FW--H analogy to reconstruct far-field hydroacoustics.
We employ third-order FVMS3 \citep{Xie2019High-fidelitysolver} and 
total variation diminishing Runge-Kutta \citep{Gottlieb1998TVD} schemes for the spatial and temporal discretization of Navier-Stokes equations where the fractional step approach \citep{Chorin1968Numerical} is used for the pressure-velocity coupling.
The first-order derivatives are approximated from quadratic polynomial and linear interpolation for the internal cells and wall boundaries respectively. The numerical solver is developed by merging the templates provided by the OpenFOAM source code. For more algorithmic details, please refer to Xie et al.~\citep{Xie2019High-fidelitysolver,Xie2020consistent}.
Figure~\ref{fig:flow_chart_hybrid_method} gives a flow chart for the whole solution procedure of the hybrid method in a time step. The solution procedure for the hybrid method is expressed in the semi-discrete formulations as follows.

\begin{enumerate}[Step (1)]
\item Given the velocity $\widetilde{u}_i^n$, and the effective eddy-viscosity $\nu_{\mathrm{eff}}^n$ at time level $t^n$, the intermediate velocity $\widetilde{u}_i^\ast$ is obtained by computing the convection and diffusion terms of the momentum equation:

\begin{equation}
	\frac{\partial \widetilde{u}_i}{\partial t}=-\frac{\partial \widetilde{u}_i \widetilde{u}_j}{\partial x_j}+\nu_{\mathrm{eff}}^n \frac{\partial^2 \widetilde{u}_i}{\partial x_j \partial x_j}.\label{eq-adv-diff}
\end{equation}
{\color{black}It is noted that the treatment of velocity boundary condition is consistent with the work in \citet{Kim1985Application,LE1991369}.}

\item The pressure Poisson equation is constructed to compute the pressure field $\widetilde{p}^{n+1}$ by enforcing the divergence-free condition,
\begin{equation}
\frac{\partial^2 \widetilde{p}^{n+1}}{\partial x_i \partial x_i} = \frac{1}{\Delta t} \frac{\partial \widetilde{u}_i^\ast}{\partial x_i}.\label{eq-poiss}
\end{equation}

\item The velocity field is updated by the projection step to satisfy the continuity equation. The velocity $\widetilde{u}_i^{n+1}$ satisfying the divergence condition is utilized to calculate the transport velocity in the subsequent time iteration. 
\begin{equation}
    \widetilde{u}_i^{n+1} = \widetilde{u}_i^\ast-\Delta t \frac{\partial \widetilde{p}^{n+1}}{\partial x_i}.
\end{equation}

\item Update the eddy-viscosity to the intermediate quantity $\nu_{\mathrm{eff}}^\ast$ at the LES grids by the WALE model.

\item Calculation of time-averaged quantities and start LES-to-WM communication: 
The time-averaged velocity $\langle \widetilde{u}_i^{n+1} \rangle$ and pressure $\langle \widetilde{p}^{n+1} \rangle$ are computed at the matching height $h_{wm}$ instead of instantaneous velocity and pressure. \textcolor{black}{Note that the time-averaging period is specified with 20 time steps for the SUBOFF.} The LES-to-WM communication is initiated, and the wall shear stress $\langle \tau_w^{n+1} \rangle$ is calculated using Eq.~\eqref{eq:wall_shear_stress}.

\item Start the WM-to-LES communication and update the wall effective eddy-viscosity at the LES grids: The WM-to-LES communication is started, and the wall shear stress $\langle \tau_w^{n+1} \rangle$ is used to update the wall effective eddy-viscosity $\nu_{\mathrm{eff}}^{n+1}$ by Eq.~\eqref{eq:UpdateNuEff} at time level $t^{n+1}$.

\item The FW--H acoustic analogy postprocessing is initiated. The direction \(\boldsymbol{r}/|\boldsymbol{r}|\) and distance \(|\boldsymbol{r}|\) of each face on each FW--H control surface are initialized for each observer. The contribution of each wall cell towards the dipole or loading component is then calculated in the FW--H equation by Eq.~\eqref{eq:FWH_equation}.

\item Integration is performed for far-field sound pressure by summing the contributions for acoustics from all surface cells for each observer to obtain the far-field sound pressure $\hat{p}^{n+1}$ at time level $t^{n+1}$.

\item Proceeding to the next time level: Subsequently, the simulation proceeds to the next time level, advancing from $t = n$ to $t = n + 1$.
\end{enumerate}

\section{Numerical validation for turbulent flows pase a cylinder}\label{numerical_validation}

To assess the capabilities of the present model in predicting hydrodynamic noise from turbulent flows, we first carry out numerical simulation for the turbulent noise of flow past a cylinder at a subcritical Reynolds number of $\mathrm{Re}=1.0\times10^4$ which provides experimental measurements and reference solutions for both the turbulent flow and acoustic field.
Since the laminar boundary layer near the wall can be directly resolved by relatively coarse LES grids, so we do not use wall modeling for this problem. Although the near-wall region in the cylinder case is fully resolved, it provides a baseline for our LES with a high-order scheme and FW--H methodology.

\begin{figure}[htbp]
	\centering
	\begin{overpic}[width=11cm]{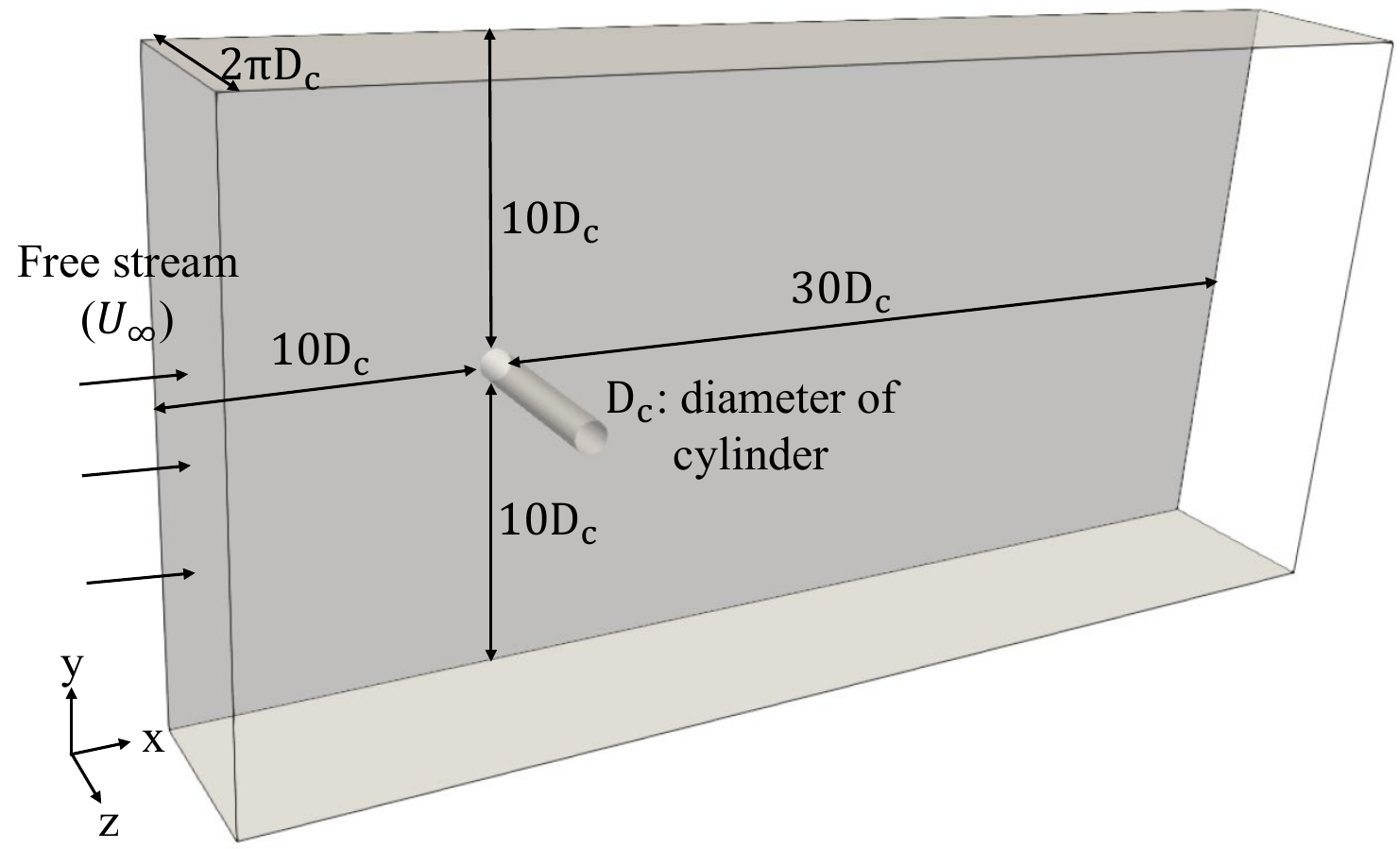}
	\end{overpic}		
	\caption[]{Computational domain used to simulate flow around a cylinder.}
	\label{fig:cylinder_computational_domain}
\end{figure}

\subsection{Computational setup}\label{cylinder_setup}

The computation setup is illustrated in Fig.~\ref{fig:cylinder_computational_domain} which is consistent with the work in \citet{Chen2023Cylinder}. 
A circular cylinder with a diameter \(D_c\) is positioned on the vertical symmetry plane of a hexahedral domain, situated at a distance of \(10D_c\) from the inlet boundary and \(30D_c\) from the downstream exit plane where the subscript `$c$' denotes the cylinder. The spanwise length of the cylinder is \(2\pi D_c\), and the vertical distance from the cylinder to each of the upper and lower sidewalls is \(10D_c\), deemed sufficiently large to avoid undesirable numerical constraints from the sidewalls. The Reynolds number $\mathrm{Re}$ is computed with the cylinder diameter \(D_c\), inflow velocity \(U_{\infty}\) and kinematic viscosity \(\nu\).

For the boundary conditions, we specify a uniform flow velocity with a fixed \(U_{\infty}\) at the inlet, an inlet-outlet boundary condition at the downstream plane to control undesired backflow and a non-slip boundary condition on the cylinder. Symmetry conditions are applied to the top, bottom, and side boundaries. A boundary layer mesh is strategically implemented around the cylinder to capture boundary layer flows. Regarding near-wall resolution, the dimensionless wall-normal length \(y^{+}\) is limited to less than 0.8 by controlling the first-layer heights (\(y^{+}=y u^{*} / \nu\), with \(u^{*}\) denoting the friction velocity). The length in the spanwise direction \(\delta z\) of a mesh cell is \(0.025D_c\), and the total mesh count is approximately 16.6 million. To ensure iterative convergence in solving PDEs, the Courant number \(Co\) is restricted to less than 0.8 throughout all simulations. Following the initial transient stages, time series data of fluid properties are collected over 80 shedding cycles for subsequent analysis. The simulations and statistical data collection are executed on a 1280-core computer cluster, consuming approximately 44\,000 CPU hours.

\subsection{Flow characteristics and far-field noise}\label{cylinder_results}

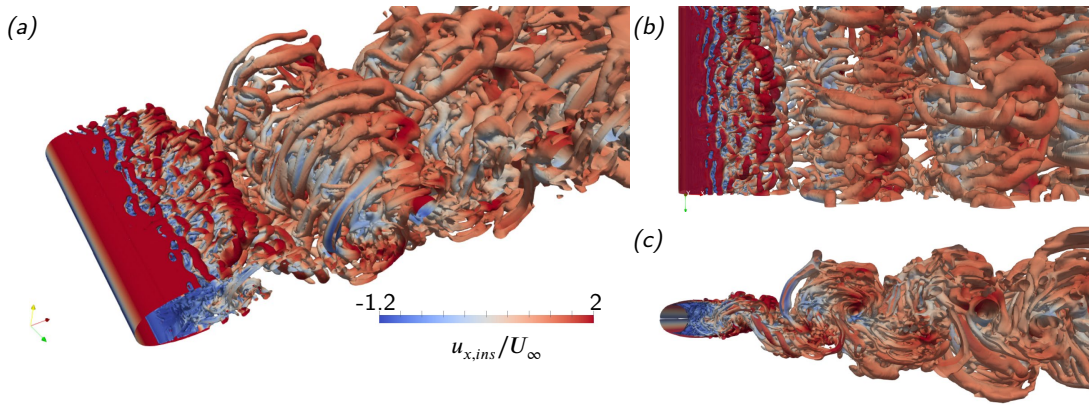
\begin{figure}[htbp]
	\centering
	\begin{overpic}[width=14cm]{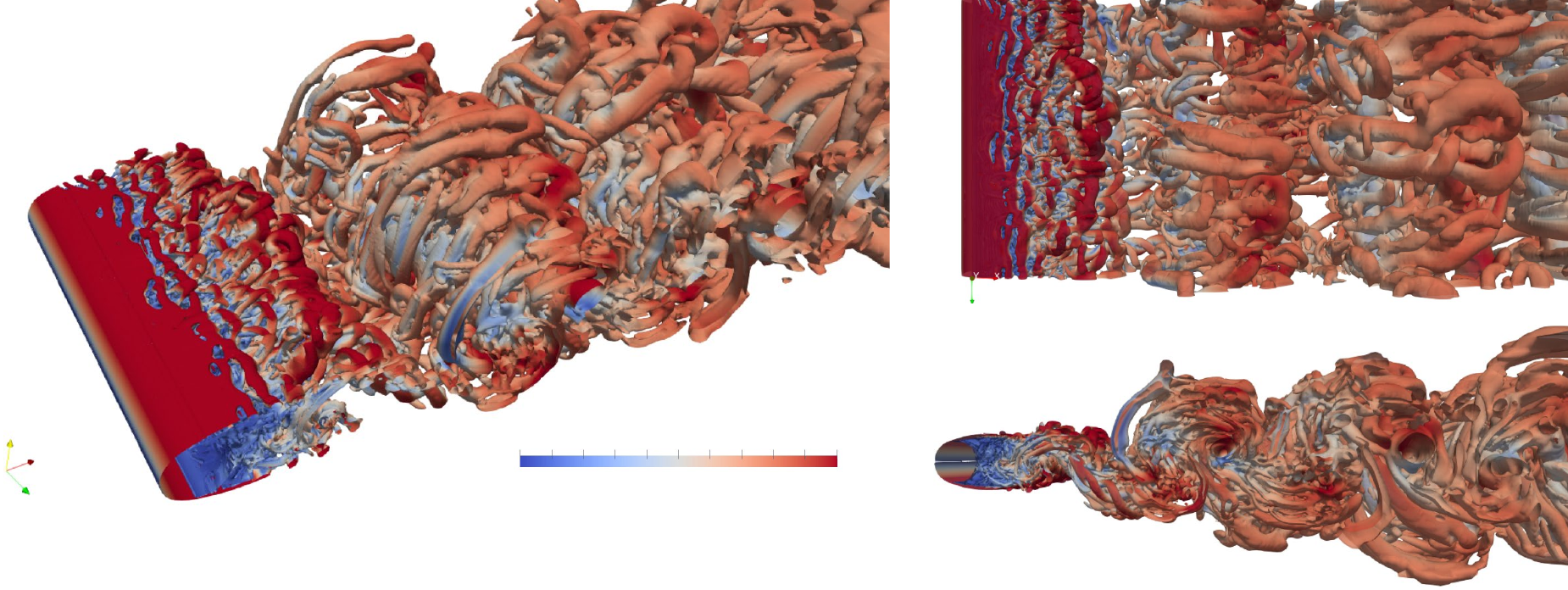}
    \put(-2,35){\color{black}{\textit{(a)}}}
    \put(57,35){\color{black}{\textit{(b)}}}
    \put(57,15){\color{black}{\textit{(c)}}}
    \put(40,5){\color{black}{$u_{x,ins}/U_{\infty}$}}
    \put(31,9){\color{black}{-1.2}}
    \put(53,9){\color{black}{2}}
	\end{overpic}		
	\caption[]{The vortex structures in turbulent flow around a cylinder at $\mathrm{Re}=1.0\times10^4$ are visualized using the iso-surface of the \textit{Q} criterion, where $QD_c^2/U_{\infty}^2 = 7\times10^{-4}$ at $t^* = 600$ ($t^* = t U_\infty /D_c$). Coherent structures are color-coded based on the instantaneous streamwise velocity, as shown in (a) isometric view, (b) top view, and (c) front view. }
	\label{fig:cylinder_vortex_structures}
\end{figure}

In Fig.~\ref{fig:cylinder_vortex_structures}, the instantaneous vortex structures around a cylinder at $\mathrm{Re}=1.0\times10^4$ are depicted, identified by the \textit{Q}-criterion proposed by \citet{hunt1988eddies}. As the flow passes through the cylinder at a subcritical Reynolds number, the boundary flow undergoes a transition from a laminar to a turbulent state, characterized by the unsteady shedding of vortex structures. Additionally, hairpin-like structures are observed just beyond the recirculation region, where the vortices alternately separate into the wake. The present high-order model accurately predicts the vortex leg, head of the shedding hairpin, and shear layer instability, as shown in Fig.~\ref{fig:cylinder_vortex_structures}. 

\begin{table}[width=.9\linewidth,cols=4,pos=htbp]
\caption{Comparison of statistical parameters for flow past a cylinder at $\mathrm{Re}=1.0\times10^4$ with available references. The statistics are the mean drag coefficient $C_{D,mean} = F_D/(\frac{1}{2}\rho U^2_\infty A)$, the Strouhal number of vortex shedding $\mathrm{St} = f_{vs}D/U_\infty$, and the root mean square of lift coefficient $C_{L,rms} = F_L/(\frac{1}{2}\rho U^2_\infty A)$}
\label{tab:cylinder_force}
\begin{tabular*}{\tblwidth}{@{} LLLL@{} }
\toprule
Case & $C_{D,mean}$ & $\mathrm{St}$ & $C_{L,rms}$\\
\midrule
Present-LES & 1.181 & 0.205 & 0.382 \\
\citet{Gopalkrishnan1993DNS} -DNS & 1.186 & 0.193 & 0.384 \\
\citet{Dong2005DNS} -DNS & 1.143 & 0.203 & 0.448 \\
\bottomrule
\end{tabular*}
\end{table}
 
Table~\ref{tab:cylinder_force} presents some statistical parameters of turbuelent flows, demonstrating that the results of $C_{D,mean}$ and $\mathrm{St}$ closely match those reported in previous DNS studies. Furthermore, the angular distribution of the mean pressure coefficient ($C_p = ({p-p_\infty})/{\frac{1}{2} \rho {U_{\infty}}^2}$) from $\theta = 0^\circ$ to $\theta = 180^\circ$ are depicted in Fig.~\ref{fig:cylinder_Cp} which shows excellent agreement with the  DNS data of \citet{Dong2005DNS} based on the spectral element framework. Quantitatively, taking the DNS data \citep{Dong2005DNS} as a reference, The maximum relative error between our LES calculations and their results is 4.11\%, which shows distinct improvements against the results in the previous study.

\begin{figure}[htbp]
	\centering
	\begin{overpic}[width=7cm]{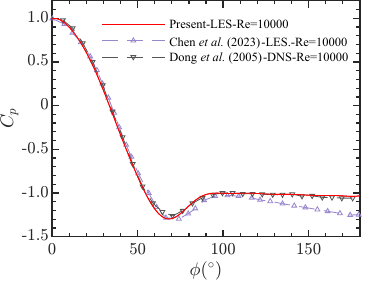}
    \put(70,35){\includegraphics[scale=0.15]{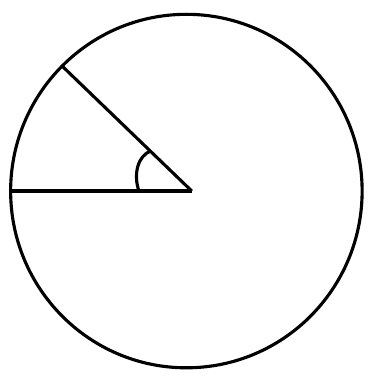}}
    \put(65,40){\footnotesize\color{black}{$0^\circ$}}
    \put(71,42){\footnotesize\color{black}{$\theta$}}
    \put(84,40){\footnotesize\color{black}{$180^\circ$}}
	\end{overpic}		
	\caption[]{\textcolor{black}{Angular distribution of mean pressure coefficient around the cylinder at $\mathrm{Re}=1.0\times10^4$ compared to LES results from \citet{Chen2023Cylinder}, DNS results from \citet{Dong2005DNS}. Reference to the angular distribution of the cylinder is given in the insets.}}
	\label{fig:cylinder_Cp}
\end{figure}
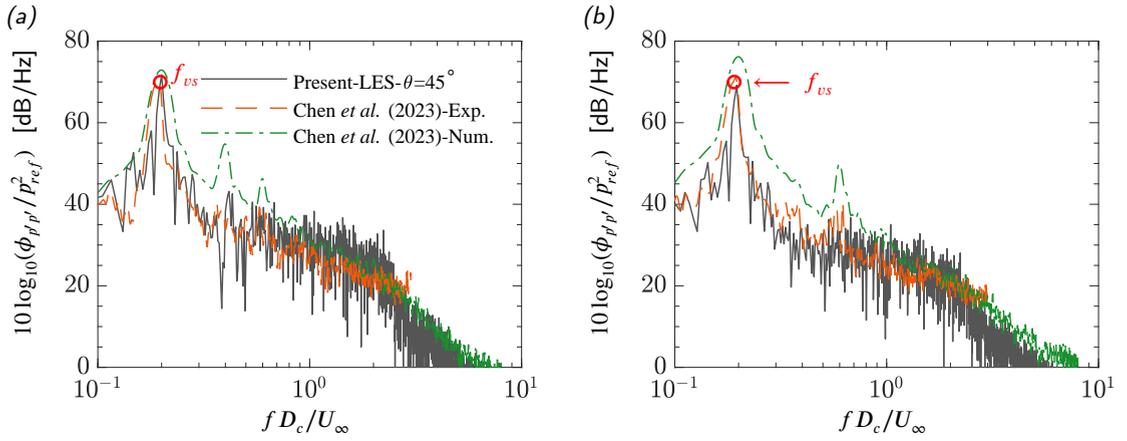
\begin{figure}[htbp]
	\centering
	\begin{minipage}[c]{1\linewidth}
		\centering
		\begin{overpic}[width=7.5cm]{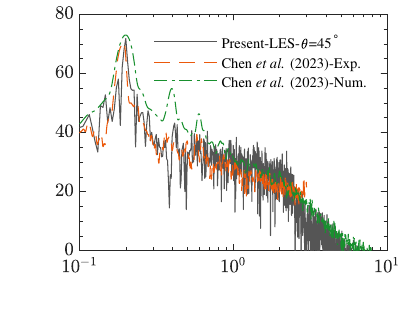}
			\put(3,75){\color{black}{\textit{(a)}}}
            \put(4,20){\color{black}{\rotatebox{90}{$10\log _{10}(\phi_{p^\prime p^\prime}/p^2_{ref})$ \  [dB/Hz]}}}
            \put(48,4){\color{black}{$fD_c/U_\infty$}}
            \put(28.5,63){\Large{\color{red}{$\circ$}}}
            \put(32,65){{\color{red}{$f_{vs}$}}}
		\end{overpic}
		\begin{overpic}[width=7.5cm]{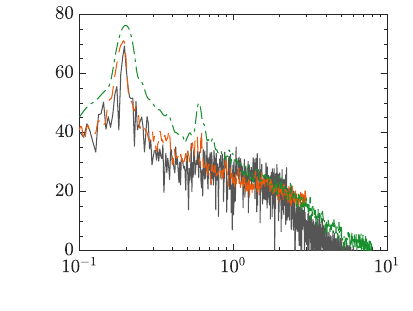}
			\put(3,75){\color{black}{\textit{(b)}}}
            \put(4,20){\color{black}{\rotatebox{90}{$10\log _{10}(\phi_{p^\prime p^\prime}/p^2_{ref})$ \  [dB/Hz]}}}
            \put(48,4){\color{black}{$fD_c/U_\infty$}}
            \put(28,63){\Large{\color{red}{$\circ$}}}
            \put(33,63.5){{\color{red}{$\longleftarrow$}}}
            \put(42,63.5){{\color{red}{$f_{vs}$}}}
		\end{overpic}
	\end{minipage}
	\caption[]{Spanwise length-velocity corrected power spectrum density of far-field acoustic pressure for flow past a cylinder at $\mathrm{Re}=1.0\times10^4$ at (a) $\theta=45^\circ$ and (b) $\theta=90^\circ$ compared to experimental and numerical results from \citet{Chen2023Cylinder}. The red circle denotes the primary peak at the vortex shedding frequency ($f_{vs}$).}
	\label{fig:cylinder_power_Spectrum}
\end{figure}

To make comparisons with the experimental results in \citet{Chen2023Cylinder}, the far-field sound pressure is sampled on a circle with a radius of $100D_c$ from the cylinder center at $z/D_c = 0$, covering $\theta = 0^\circ$ to $\theta = 360^\circ$ with a $10^\circ$ increment. Welch's spectrum distributions \citep{Welch1967} of the sound pressure at various measurement locations are analyzed. As the wind tunnel experiments have different setups from the present simulations, the simulated results are corrected for cylinder span length and free flow velocity to facilitate comparison with the experimental results, following the correction method in \citet{kato1993numerical} and \citet{Phillips1956}, consistent with the approach used by \citet{Chen2023Cylinder}.
\begin{figure}[htbp]
	\centering
	\begin{overpic}[width=7cm]{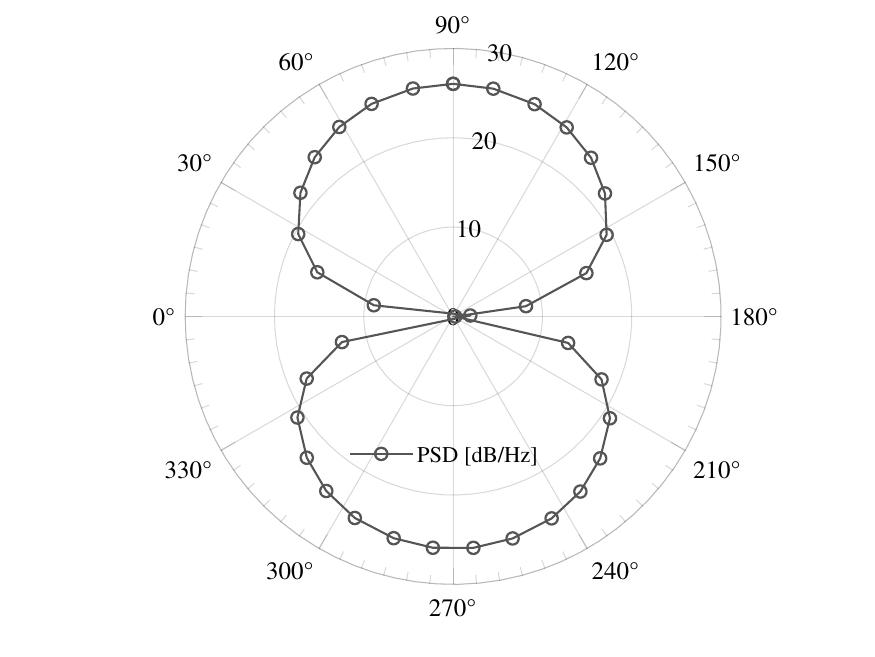}
	\end{overpic}		
	\caption[]{Far-field noise directivities calculated from the spanwise length-velocity corrected PSD at the fundamental vortex shedding frequency ($f_{vs}$).}
	\label{fig:cylinder_directivity}
\end{figure}
Figure~\ref{fig:cylinder_power_Spectrum} presents the spanwise length-velocity corrected power spectral density (PSD) of far-field acoustic pressure at $\theta = 45^\circ$ and $\theta = 90^\circ$, respectively. The primary peak at the vortex shedding frequency ($f_{vs}$) and the slope of the spectrum in the present simulation closely match the reference experimental results reported in \citet{Chen2023Cylinder}. However, the second harmonic at $f=2f_{vs}$ is slightly underpredicted at $\theta = 90^\circ$. Quantitatively, a sound peak at the $f_{vs}$ is observed with a deviation of approximately 0.88 dB/Hz and -1.65 dB/Hz at $\theta = 45^\circ$ and $\theta = 90^\circ$,  respectively. Figure~\ref{fig:cylinder_directivity} illustrates the sound directivities at the shedding frequency. The \textit{SPL} directivity displays a typical dipole pattern with symmetric characteristics. 
The results indicate that the present model accurately reproduces the principal flow characteristics and sound propagation mechanisms, establishing the groundwork for the upcoming investigation of turbulent noise of flow over an axisymmetric hull.

\section{Hydrodynamic noise of turbulent flow over an axisymmetric hull}\label{suboff}
\subsection{Computational setup and hydroacoustic post-processing
}\label{suboff_simulation_setup}
\begin{figure}[htbp]
	\centering
	\begin{overpic}[width=13cm]{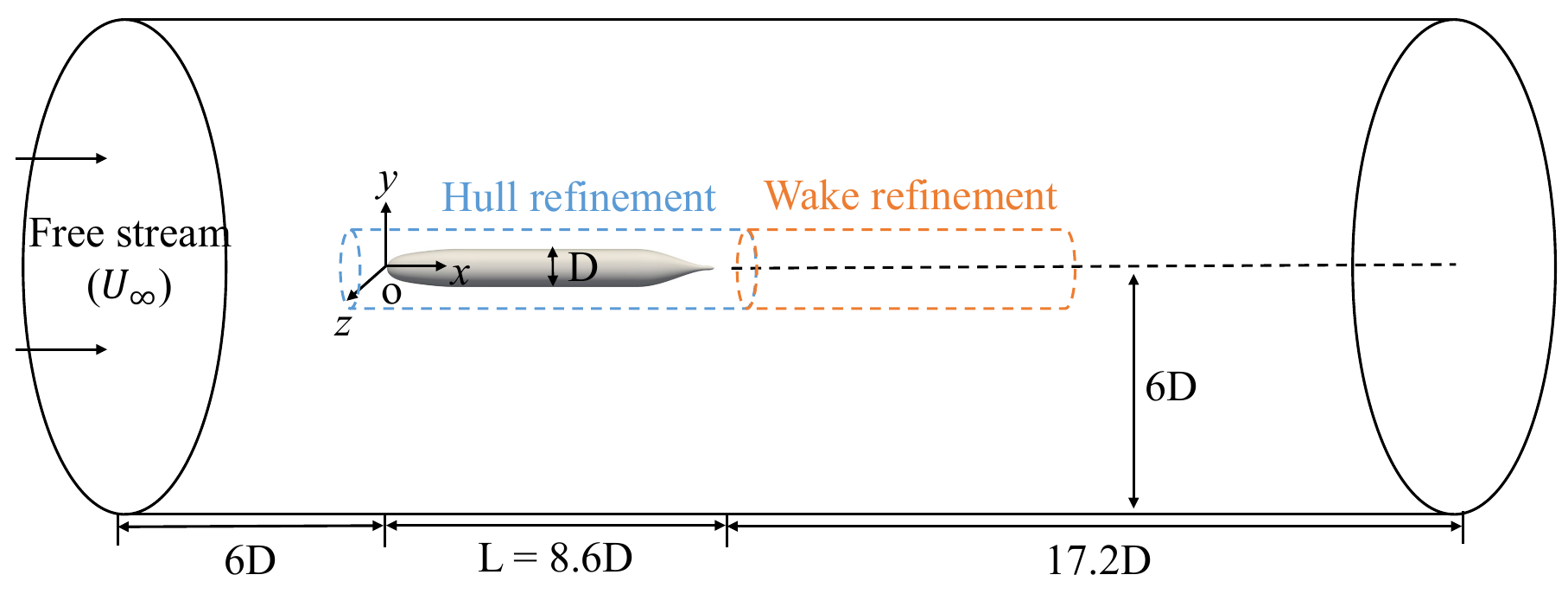}
	\end{overpic}		
	\caption[]{Computational domain and boundary conditions for DARPA SUBOFF simulation. Grid refinement is applied in the near-wall and wake regions which are indicated in blue and orange colors.}
	\label{fig:suboff_computational_domain}
\end{figure}
This section aims to investigate turbulent flow and associated far-field turbulent noise for flow past the SUBOFF hull \citep{groves1989geometric}. The numerical setup is illustrated in Fig.~\ref{fig:suboff_computational_domain} which follows the configurations of \citet{Kumar2018SUBOFF} and \citet{morse2021suboff}. As shown in Fig.~\ref{fig:suboff_computational_domain}. The computational domain is cylindrical which spans $28.8D$ in length and has a radius of $6D$. A SUBOFF model is placed at the origin of the coordinate with the maximum hull diameter of D. To reduce the numerical influence of inflow, the inlet boundary is placed $6D$ from the hull's front, and the domain extends $17.2D$ downstream from the hull's end. The Reynolds number, defined as $\mathrm{Re} = U_{\infty}L/\nu=1.2\times10^7$, based on the length  of SUBOFF hull ($L = 8.6D$), free stream velocity $U_{\infty}$, and kinematic viscosity $\nu$. Regarding boundary conditions, a uniform flow velocity with a fixed $U_{\infty}$ value is imposed at the inlet. Convective conditions are used for velocity, employing $U_\infty$ as the convective velocity, to regulate nonphysical backflow at the outlet boundary. We specify the Neumann boundary conditions for the velocity at the lateral boundary and the non-slip boundary condition
for the velocity at the hull. Neumann boundary conditions are applied for pressure and viscosity at all computational domain boundaries. 

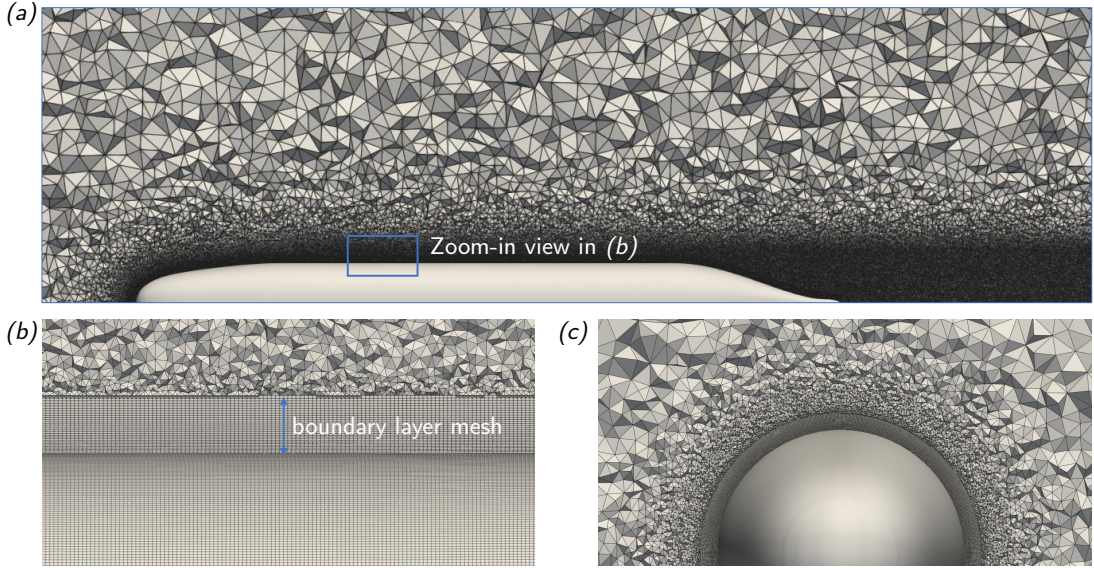
\begin{figure}[htbp]
	\centering
	\begin{overpic}[width=14cm]{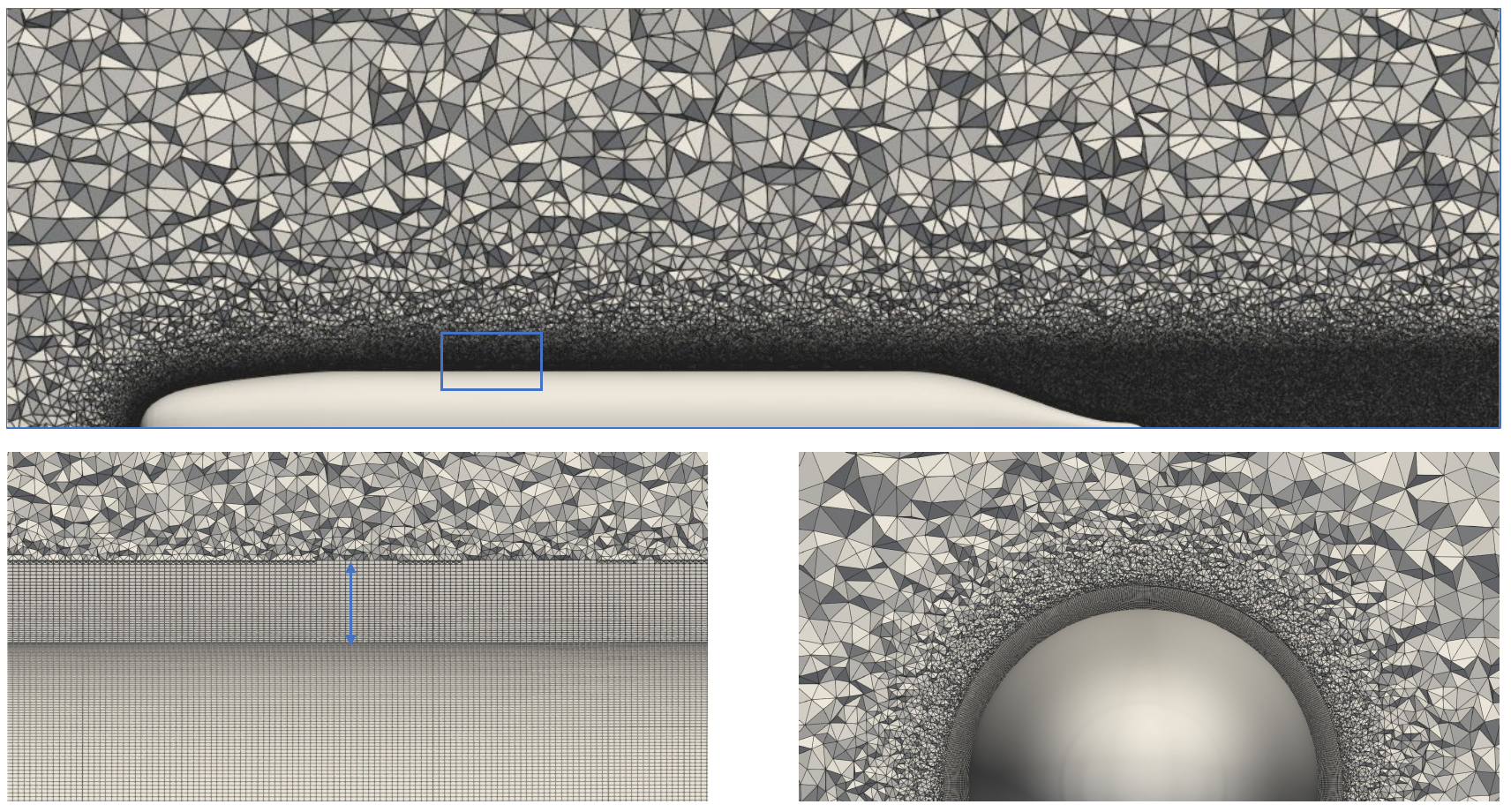}
    \put(-3,52){\color{black}{\textit{(a)}}}
    \put(37,30){\color{white}{Zoom-in view in \textit{(b)}}}
    \put(-3,22){\color{black}{\textit{(b)}}}
    \put(24,13){\color{white}{boundary layer mesh}}
    \put(49,22){\color{black}{\textit{(c)}}}
	\end{overpic}		
	\caption[]{Details of the computational grids used for DARPA SUBOFF simulation: (a) Side view of the refined grid around the hull; (b) Enlarged view of the boundary layer mesh; and (c) Front view of the unstructured grid in the vicinity of the bow.}
	\label{fig:suboff_mesh}
\end{figure}

The computational domain is divided into hybrid unstructured grids which are shown in Fig.~\ref{fig:suboff_mesh} for the mesh structure around the hull. We refine the computational grids near the hull walls and wake zones within a cylindrical region of diameter of $D+4\delta$ where $\delta$ is the thickness of the TBL at the end of the central parallel body. The `hull refinement' region, connected to the hull, spans from $0.5D$ in front of the nose to $0.5D$ behind the end of the hull and the `wake refinement' region, sharing the same radius, extends from $0.5D$ to $5D$ behind the end of the hull. Grid stretching is applied downstream and towards the lateral cylindrical boundary. The first wall-normal grid thickness is $0.02\delta$ with a growth ratio of 1.025. It ensures approximately 33 cells across the boundary layer thickness which meets the requirement of mesh resolution of WMLES. This results in a more refined wall-parallel grid resolution ($\Delta x = \Delta z = \delta/20$)
for the prediction of turbulent noise. As mentioned in 
\citet{Park2016wallPressure}, higher grid resolution is required to predict the near-wall pressure fluctuations
which are important acoustic source \citep{Wang2006noiseReview}. In total, the grid comprises 89 million control volumes across the entire computational domain. The increased grid size aims to refine the hull surface for precise prediction of surface pressure fluctuations. \textcolor{black}{Notably, in the present study, the matching height  \(h_{wm}\) is set to the centroids of the fifth cells off the hull from the LES grids to achieve a good accuracy of the near-wall pressure fluctuations, resulting in $h_{wm}\approx 0.1\delta$ or $h^+_{wm} \approx$ 250 in inner units. This choice was made following the suggestion
 from previous research \cite{Kawai2012WallmodelingIL,Boukharfane2020WallPressure,Hu2023WMLES}.}

To reproduce the experimental conditions of \citet{Huang1992Exp}, the boundary layer is tripped by imposing a steady wall-normal velocity of $0.06U_\infty$ at this $x/D = 0.75$. This tripping method effectively accelerates the laminar-to-turbulent transition which is in alignment with the numerical and experimental setup in prior investigations \citep{Kumar2018SUBOFF, morse2021suboff}. To improve the numerical efficiency,
we accelerate the flow development in the present simulation which are segmented into three stages: (i) The initial development of the flow around the SUBOFF hull occurs at a lower $\mathrm{Re} = 1.2\times10^6$ using RANS simulations (employing the $k-\omega$ SST model) without any boundary layer perturbation (no trip wire); (ii) The Reynolds number is then increased to $\mathrm{Re}=1. 2\times10^7$ through LES without wall models for one flow-through time to simulate an initial field for WMLES, and the boundary layer is tripped. \textcolor{black}{Note that a flow-through time is taken as the duration in which the freestream flows through entire the hull (L/$U_\infty$), where $L$ is the hull length and $U_\infty$  is the free-stream velocity;} (iii) Subsequently, WMLES of the flow past the hull at $\mathrm{Re} = 1.2\times10^7$ with boundary tripping is computed for two flow-through times, ensuring the flow is free of initial transients and considered fully developed. Flow statistics are collected over two further flow-through times through temporal and azimuthal averaging, which is consistent with the approach used in \citet{posa2016suboff, posa2020suboff, Chen2023SUBOFF}. The simulation is carried out with a time step of  $\Delta t U_\infty/D = 3.48\times10^{-4}$  on 1280 processors at the Center for HPC Center of Shanghai Jiao Tong University, utilizing approximately 500\,000 CPU-hours. \textcolor{black}{We also evaluate the computational cost by comparing it with a conventional scheme. The present model takes approximately twice the computational time as the conventional one on the same grid resolution which is computationally efficient if considering the improvements in numerical accuracy and solution quality. 
For more details, please refer to \citet{Xie2019High-fidelitysolver}.}

The instantaneous numerical solutions generated through WMLES computations are utilized to predict the acoustic results of the SUBOFF hull. Here, the same approach was carefully validated for the far-field noise of a cylinder in Section~\ref{numerical_validation}, where successful comparisons with wind tunnel acoustic measurements were reported. Hydrophones were positioned on a sphere with a radial extent of $500D$, roughly 58 times the hull's length, and centered on the reference frame's origin. The selection of $500D$ aligns with the far-field source region condition defined by \citet{Wang2006noiseReview} and is consistent with the value employed by \citet{Zhou2022suboff}. Employing an angular spacing of $10^\circ$ in the longitude and latitude of the spherical domain, a total of 648 uniformly distributed hydrophones were deployed to characterize far-field sound from the turbulent flow. Numerical solutions were stored at each time step ($\Delta t U_\infty/D = 3.48\times10^{-4}$), corresponding to a sampling frequency of $f_s D/U_\infty = 2\,871$. Additionally, key parameters used for calculating \textit{SPLs} include the reference pressure $\widehat{p}_{0w} = 1\ \mathrm{\mu Pa}$, water density $\rho_w = 1\,000\ \mathrm{kg/m^3}$, and speed of sound in water $c_w = 1\,482\ \mathrm{m/s}$, here, the subscript `$w$' represents that the fluid is water.

\subsection{Numerical results and discussion}\label{suboff_Results_discussion}
\subsubsection{Flow features and flow validation}\label{suboff_flow_field}
\begin{figure}[!t]
	\centering
	\begin{overpic}[width=14cm]{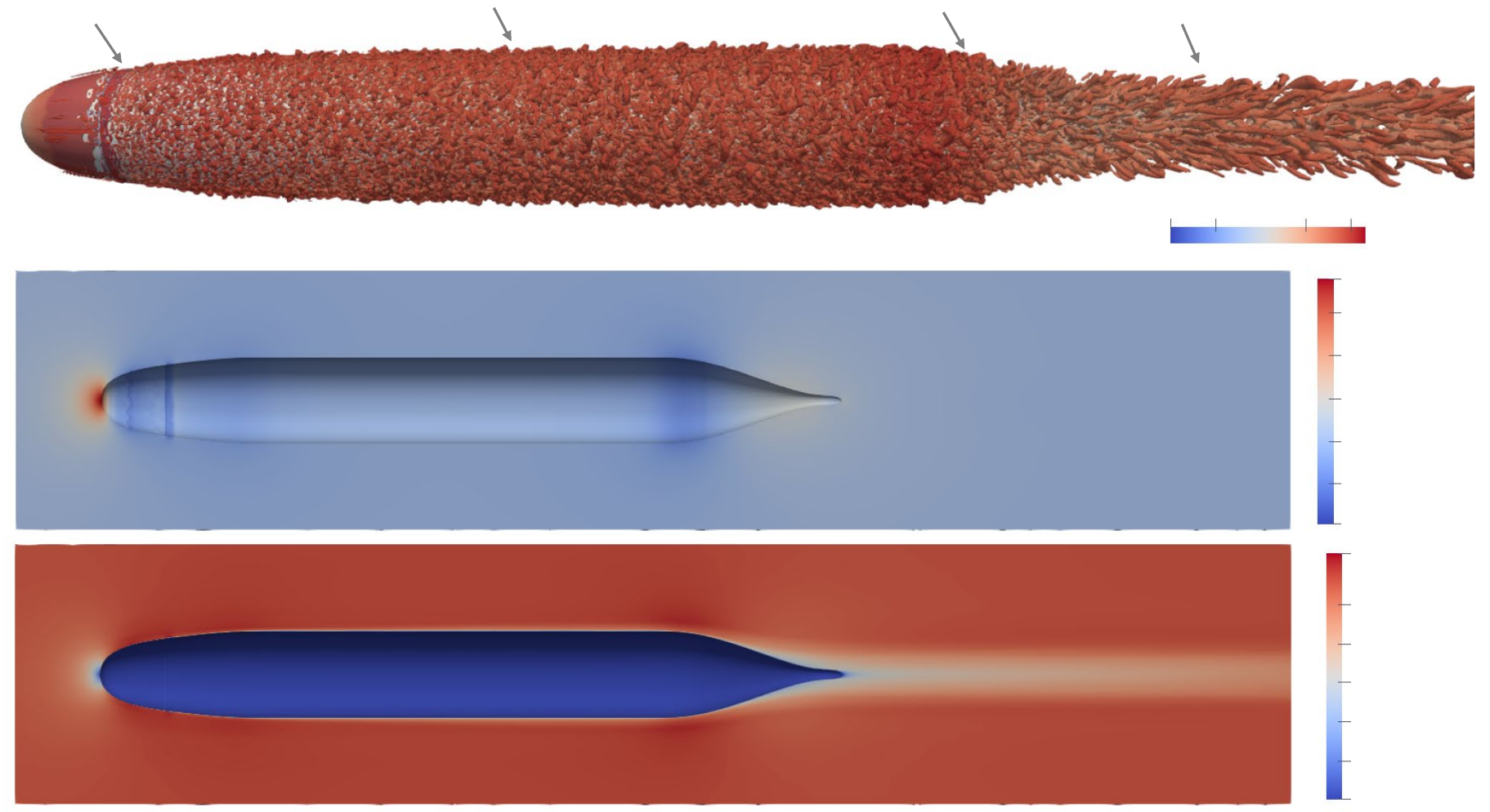}
    \put(-3,52){\color{black}{\textit{(a)}}}
    \put(0,54){\footnotesize\color{black}{Tripping location}}
    \put(20,55){\footnotesize\color{black}{Turbulent boundary layer}}
    \put(48,55){\footnotesize\color{black}{Adverse pressure gradient}}
    \put(75,54){\footnotesize\color{black}{Turbulent wake}}
    \put(81,40){\footnotesize\color{black}{$u_{x,ins}/U_{\infty}$}}
    \put(73,38){\footnotesize\color{black}{-0.36}}
    \put(92,38){\footnotesize\color{black}{1.16}}
    
    \put(-3,35){\color{black}{\textit{(b)}}}
    \put(92,27){\footnotesize\color{black}{$C_p$}}
    \put(90,35){\footnotesize\color{black}{1.0}}
    \put(91,19){\footnotesize\color{black}{-0.3}}
    
    \put(-3,16){\color{black}{\textit{(c)}}}
    \put(92,9){\footnotesize\color{black}{$|U|/U_\infty$}}
    \put(91,16){\footnotesize\color{black}{1.12}}
    \put(91,0){\footnotesize\color{black}{0}}
    
	\end{overpic}		
	\caption[]{(a) Instantaneous vortex structures using the iso-surface of the \textit{Q} criterion with $QD^2/U_{\infty}^2 = 7\times10^{-4}$ at $t^* = 1.6s$ colored by the instantaneous streamwise velocity; (b) mean pressure coefficient; and (c) mean velocity magnitude.}
	\label{fig:suboff_flow_field}
\end{figure}
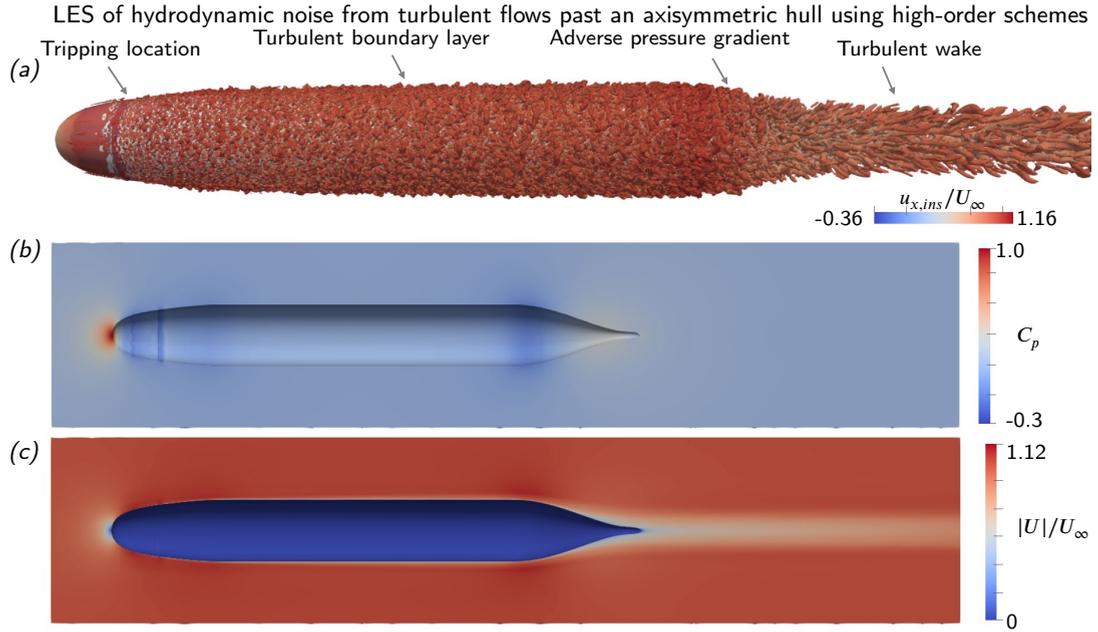

\begin{figure}[!t]
	\centering
	\begin{minipage}[c]{1\linewidth}
		\centering
		\begin{overpic}[width=8cm]{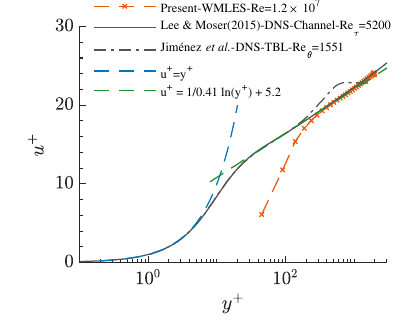}
        \put(74.5,39){\large{\color{black}{\rotatebox{90}{$\longrightarrow$}}}}
        \put(74.5,35){\color{black}{$h_{wm}$}}
		\end{overpic}
	\end{minipage}
	\caption[]{\textcolor{black}{Statistics in wall units for hull boundary layer at $x/L=0.5$ on the hull, compared to DNS results of fully developed channel flow from \citet{Lee2015Channel} and DNS of a planar TBL at
$Re_{\theta}=1551$ from \citet{TBL2010}.}}
	\label{fig:suboff_boundary_velocity_profile}
\end{figure}

The instantaneous vortex structures using the iso-surface of the \textit{Q} criterion are displayed in 
Figure~\ref{fig:suboff_flow_field} which also presents the contours of the mean pressure coefficients ($C_p$) and mean velocity magnitude ($|U|/U_\infty$). Examining the boundary layer flow, no coherent structures are observed near the fuselage nose just before the trigger point. However, the flow finally grows into a fully developed turbulent state downstream of the trigger point, as confirmed by the velocity profile at $x/L = 0.5$. In Fig.~\ref{fig:suboff_boundary_velocity_profile}, it is evident that the mean velocity profiles obtained from the present calculations at the station ($x/L = 0.5$) exhibit excellent agreement with the well-established wall law characteristic of a 2-D TBL, following $u^+ = 1/0.41 \ln{y^+} + 5.2$ in the log layer of the TBL. Furthermore, the present results exhibit a good comparison with a DNS of a channel flow at a comparable Reynolds number of $\mathrm{Re_{\tau}}$ = 5200, as investigated by \citet{Lee2015Channel}. \textcolor{black}{In Fig.~\ref{fig:suboff_boundary_velocity_profile}, the DNS results for the TBL of a flat plate at \(Re_{\theta} = 1551\), as reported by \citet{TBL2010} also used for comparison. Despite comparable boundary layer thicknesses at this Reynolds number, the hull boundary layer exhibits a greater friction velocity (\(u_{\tau}\)), resulting in reduced \(U^{+}\) values within the outer region of the turbulent layer relative to planar TBL measurements at equivalent \(Re_{\theta}\). This discrepancy primarily arises from the influence of the hull's transverse curvature on the axisymmetric TBL, a phenomenon previously corroborated by numerical investigations, such as those conducted by \citet{Kumar2018SUBOFF}.} Following the axisymmetric TBL in the mid-parallel section, the flow eventually separates at the stern, forming the wake due to the adverse pressure gradient caused by the change in hull curvature. In the wake flow further downstream, unlike the bluff-body wake observed in the turbulent wake of the cylinder in Fig.~\ref{fig:cylinder_vortex_structures}, vortex shedding does not exert dominant control over the flow in the near wake. The wake topology exhibits a quasi-cylindrical configuration, similar to observations in previous studies on axisymmetric hulls, such as those by \citet{Kumar2018SUBOFF}, \citet{morse2021suboff}, and \citet{Ortiz2021slenderBody}.

\subsubsection{Hydrodynamic force and flow characteristics on the body}\label{suboffForce}
\begin{figure}[!t]
	\centering
	\begin{minipage}[c]{1\linewidth}
		\centering
		\begin{overpic}[width=7.5cm]{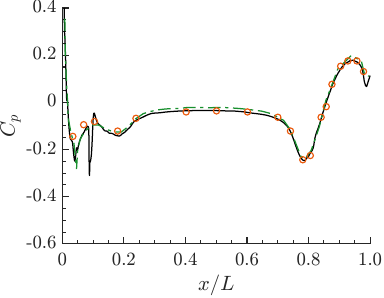}
			\put(1,75){\color{black}{\textit{(a)}}}
            \put(32,49){\footnotesize\color{red}{$P0$}}
            \put(46,51){\footnotesize\color{red}{$P1$}}
            \put(54,51){\footnotesize\color{red}{$P2$}}
            \put(62,51){\footnotesize\color{red}{$P3$}}
            \put(17,16){\includegraphics[scale=0.205]{ 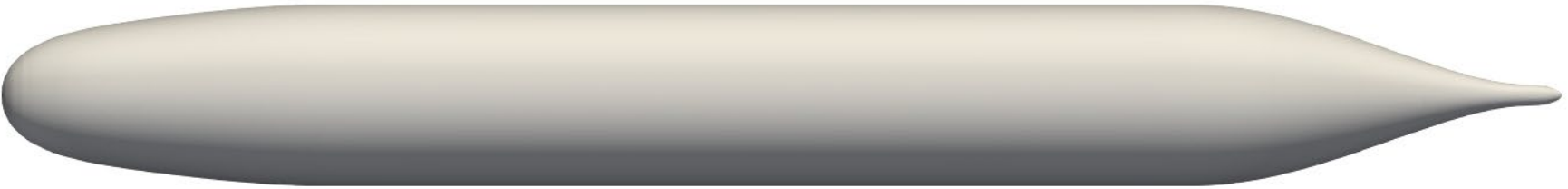}}
		\end{overpic}
		\begin{overpic}[width=7.5cm]{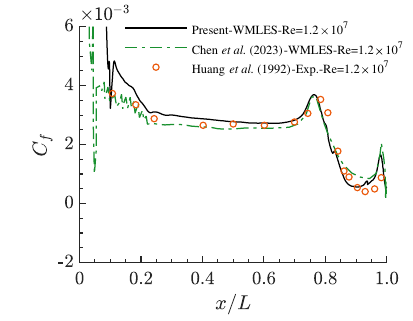}
			\put(4,75){\color{black}{\textit{(b)}}}
            \put(20,16){\includegraphics[scale=0.19]{Figure_15-1.pdf}}
		\end{overpic}
	\end{minipage}
	\caption[]{Profiles of (a) the mean surface pressure coefficients $C_p$ and (b) the mean surface skin friction coefficients $C_f$ along the hull, compared to experimental results from \citet{Huang1992Exp}, WMLES results of SUBOFF hull from \citet{Chen2023SUBOFF}.}
	\label{fig:suboff_cp_cf}
\end{figure}

Figure~\ref{fig:suboff_cp_cf} (a) demonstrates a remarkable agreement of $C_p$ along the hull length between the present results and the experimental data from \citet{Huang1992Exp}. Additionally, we report the surface friction coefficient ($C_f = \tau_w/\rho U_{\infty}^2$ with $\tau_w$ being the wall shear stress) as another classical verification parameter to assess the numerical accuracy in predicting wall shear stress. Figure~\ref{fig:suboff_cp_cf} (b) shows good agreement between our results and the experimental data in \citet{Huang1992Exp}, suggesting accurate prediction of $C_f$ along the sphere. Furthermore, it's worth noting that the present model surpasses the traditional low-order numerical models in predicting wall shear stress. We perform a further comparison for the drag coefficient, which is also related to wall shear stress. 
\begin{table}[width=.5\linewidth,cols=2,pos=htbp]
\caption{Comparison of statistical parameters on the hull surface at $\mathrm{Re}=1.2\times10^7$ with available references.}
\label{tab:suboff_force}
\begin{tabular*}{\tblwidth}{@{} LL@{} }
\toprule
Case & $C_{d,mean}$ \\
\midrule
Present-WMLES & 0.0927  \\
\citet{Huang1992Exp}-Exp. & 0.093  \\
\citet{Chen2023SUBOFF}-WMLES & 0.088  \\
\bottomrule
\end{tabular*}
\end{table}
Here the drag on the hull is denoted by $F_D$, and the reference area used to non-dimensionalize $F_D$ is defined as $A = \tfrac{1}{4}\pi D^2$. The SUBOFF hull at $\mathrm{Re} = 1.2\times10^7$ has a drag coefficient of 0.093, according to \citet{Huang1992Exp}. The drag coefficient, as shown in Table~\ref{tab:suboff_force}, is 0.0927 for the present high-order WMLES, predicted using total hull drag which closely aligns with the experimental measurements. 

\begin{figure}[!t]
	\centering
	\begin{minipage}[l]{1\linewidth}
		\centering
		\begin{overpic}[width=7.5cm]{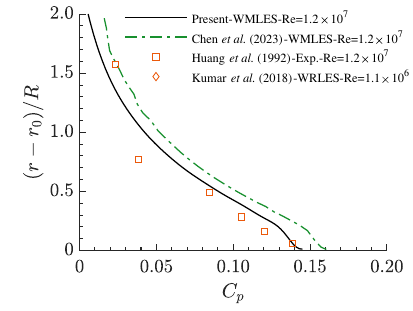}
			\put(3,70){\color{black}{\textit{(a)}}}
		\end{overpic}
		\begin{overpic}[width=7.5cm]{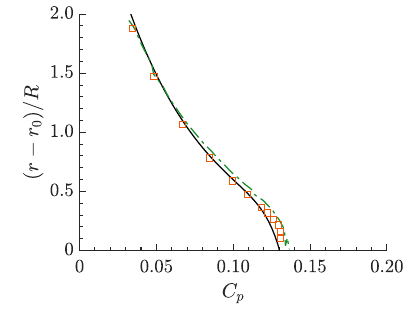}
			\put(3,70){\color{black}{\textit{(b)}}}
		\end{overpic}
	\end{minipage}
	\begin{minipage}[l]{1\linewidth}
		\centering
		\begin{overpic}[width=7.5cm]{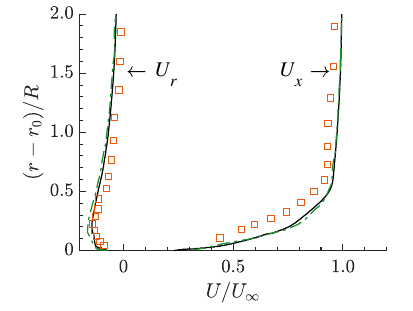}
			\put(3,70){\color{black}{\textit{(c)}}}
		\end{overpic}
		\begin{overpic}[width=7.5cm]{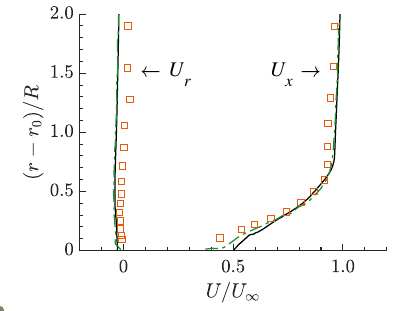}
			\put(3,70){\color{black}{\textit{(d)}}}
		\end{overpic}
	\end{minipage}
	\begin{minipage}[l]{1\linewidth}
		\centering
		\begin{overpic}[width=7.5cm]{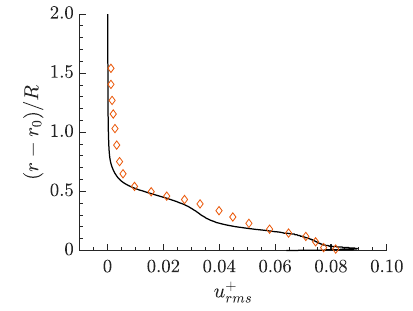}
			\put(3,70){\color{black}{\textit{(e)}}}
		\end{overpic}
		\begin{overpic}[width=7.5cm]{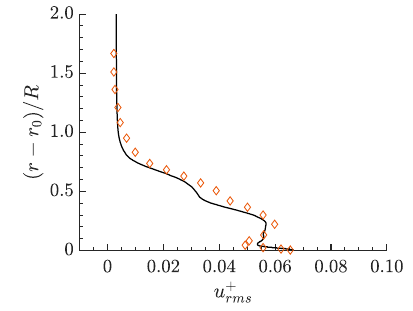}
			\put(3,70){\color{black}{\textit{(f)}}}
		\end{overpic}
	\end{minipage}
	\caption[]{Profiles of pressure coefficient ($C_p$) (a,b), mean axial ($U_x$) and radial ($U_r$) velocity (c,d), and root-mean-square of axial velocity fluctuations (e,f) at Slice-I: $x/L=0.904$ (a, c, e) and Slice-II: $x/L=0.978$ (b, d, f). Note that $r_0$ represents the local radius of the hull, while $R$ denotes the maximum radius. The comparisons include the experimental results of \citet{Huang1992Exp} and the numerical results from \citet{Chen2023SUBOFF, Kumar2018SUBOFF}.}
	\label{fig:suboff_wake_U_Cp_urms}
\end{figure}
Figure~\ref{fig:suboff_wake_U_Cp_urms} compares profiles of pressure coefficient ($C_p$), nondimensional mean axial ($U_x/U_\infty$), and radial ($U_r/U_\infty$) velocity, and root-mean-square axial velocity fluctuations at two stern locations of Slice-I at $x/L=0.904$ and Slice-II at $x/L=0.978$ where numerical and experimental results are available in \citep{Huang1992Exp, Chen2023SUBOFF, Kumar2018SUBOFF}. The mean pressure coefficient ($C_p$) and the mean velocities ($U_x,\ U_r$) obtained from the present WMLES agree well with the experimental results. 
Both results of the present model and \citet{Chen2023SUBOFF} compare well with experimental measurements of \citet{Huang1992Exp} in terms of flow statistics.
Regarding the velocity fluctuation, we can observe some minor deviation from  WRLES results of \cite{Kumar2018SUBOFF}, which may be attributed to the difference in Reynolds number of reference solutions.
It is realized that the present WMLES model provides significant advantages in numerical accuracy for solving high Reynolds number flow problems with complex geometry.

\subsubsection{Fluctuations of hydrodynamic pressure on the hull}\label{suboff_surface_wall_pressure}
\begin{figure}[!t]
	\centering
	\begin{minipage}[c]{1\linewidth}
		\centering
		\begin{overpic}[width=7.5cm]{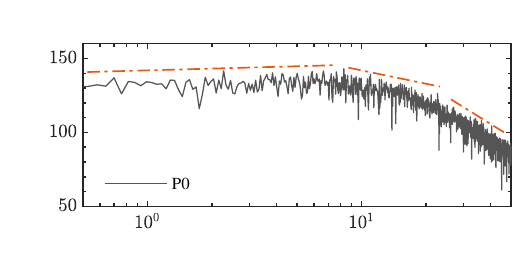}
			\put(-5,45){\color{black}{\textit{(a)}}}
            \put(30,37.5){\footnotesize{\color{red}{$\propto \mathrm{St}^{0.4}$}}}
            \put(70,37){\footnotesize{\color{red}{\rotatebox{-10}{$\propto \mathrm{St}^{-3}$}}}}
            \put(85,33){\footnotesize{\color{red}{\rotatebox{-30}{$\propto \mathrm{St}^{-9}$}}}}
            \put(2,5){\footnotesize{\color{black}{\rotatebox{90}{$10\log _{10}(\phi_{p^\prime p^\prime}/p^2_{ref})\ \mathrm{[dB/Hz]}$}}}}
            \put(48,2){\footnotesize{\color{black}{St=$fD_c/U_\infty$}}}
		\end{overpic}
  	\begin{overpic}[width=7.5cm]{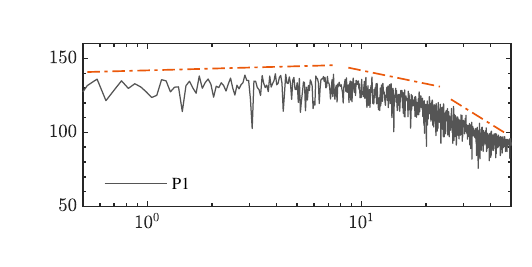}
			\put(-5,45){\color{black}{\textit{(b)}}}
            \put(30,37.5){\footnotesize{\color{red}{$\propto \mathrm{St}^{0.4}$}}}
            \put(70,37){\footnotesize{\color{red}{\rotatebox{-10}{$\propto \mathrm{St}^{-3}$}}}}
            \put(85,33){\footnotesize{\color{red}{\rotatebox{-30}{$\propto \mathrm{St}^{-9}$}}}}
            \put(2,5){\footnotesize{\color{black}{\rotatebox{90}{$10\log _{10}(\phi_{p^\prime p^\prime}/p^2_{ref})\ \mathrm{[dB/Hz]}$}}}}
            \put(48,2){\footnotesize{\color{black}{St=$fD_c/U_\infty$}}}
		\end{overpic}
    	\begin{overpic}[width=7.5cm]{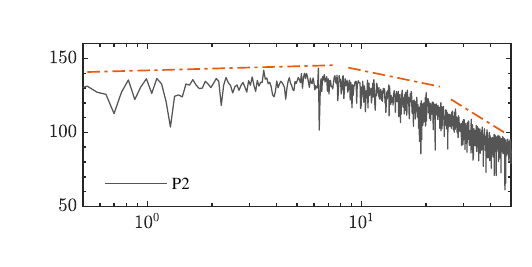}
			\put(-5,45){\color{black}{\textit{(c)}}}
            \put(30,37.5){\footnotesize{\color{red}{$\propto \mathrm{St}^{0.4}$}}}
            \put(70,37){\footnotesize{\color{red}{\rotatebox{-10}{$\propto \mathrm{St}^{-3}$}}}}
            \put(85,33){\footnotesize{\color{red}{\rotatebox{-30}{$\propto \mathrm{St}^{-9}$}}}}
            \put(2,5){\footnotesize{\color{black}{\rotatebox{90}{$10\log _{10}(\phi_{p^\prime p^\prime}/p^2_{ref})\ \mathrm{[dB/Hz]}$}}}}
            \put(48,2){\footnotesize{\color{black}{St=$fD_c/U_\infty$}}}
		\end{overpic}
    	\begin{overpic}[width=7.5cm]{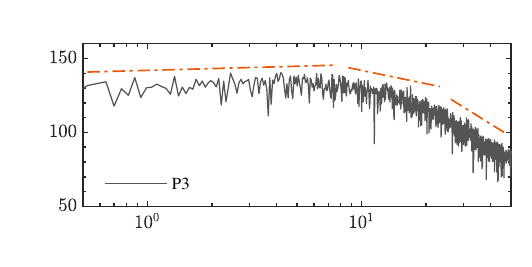}
			\put(-5,45){\color{black}{\textit{(d)}}}
            \put(30,37.5){\footnotesize{\color{red}{$\propto \mathrm{St}^{0.4}$}}}
            \put(70,37){\footnotesize{\color{red}{\rotatebox{-10}{$\propto \mathrm{St}^{-3}$}}}}
            \put(85,33){\footnotesize{\color{red}{\rotatebox{-30}{$\propto \mathrm{St}^{-9}$}}}}
            \put(2,5){\footnotesize{\color{black}{\rotatebox{90}{$10\log _{10}(\phi_{p^\prime p^\prime}/p^2_{ref})\ \mathrm{[dB/Hz]}$}}}}
            \put(48,2){\footnotesize{\color{black}{St=$fD_c/U_\infty$}}}
		\end{overpic}
	\end{minipage}
	\caption[]{Surface pressure PSD of pressure fluctuations at different locations over the parallel mid-hull (a) $P0: x/L = 0.239$ (b) $P1: x/L = 0.402$ (c) $P2: x/L = 0.501$, and (d) $P3: x/L = 0.601$. Refer to Fig.~\ref{fig:suboff_cp_cf} (a) for the location schematic of probes.}
	\label{fig:suboff_wall_pressure_fluc}
\end{figure}
The computation of hydrodynamic pressure fluctuations 
on the SUBOFF hull is crucial for the prediction of the turbulent noise in far-field \citep{Wang2006noiseReview}. 
Previous investigations \cite{Posa2023propeller, Zhou2022suboff} have highlighted the dominance of loading noise in the hydroacoustic analysis of flow past underwater blunt bodies. Such noise originates from fluctuations in hydrodynamic pressure on the bodies' surfaces. To address the subsequent discussion on the acoustic signature of the flow around the SUBOFF hull, we analyze the surface pressure PSD of pressure fluctuations at specific locations of $x/L = 0.239,\ 0.402,\ 0.501,\ 0.601$ which are depicted in Figure~\ref{fig:suboff_cp_cf} (a) along the parallel mid-hull.

The pressure signals are extracted from probes located at the first off-wall cell center in our simulation and
the power spectra of wall pressure fluctuations are estimated by the Welch method of periodogram as \citep{Welch1967}. We collect a dataset of 50\,000 pressure signals for each probe and employ FFT for the spectral analysis. The PSD is plotted against the normalized frequency, represented by the Strouhal number, defined as $\mathrm{St} = fD/U_\infty$, at various streamwise locations. In Fig.~\ref{fig:suboff_wall_pressure_fluc}, the PSD reveals a low-frequency broadband spectrum, with the majority of signal energy below $\mathrm{St} = 10$. Minor differences are observed in the PSD of probes at different locations in the parallel mid-hull. The superimposed slopes on each dimensionless PSD indicate that the spectra of wall pressure fluctuations vary roughly as $\mathrm{St}^{-9}$ at high frequencies, $\mathrm{St}^{-3}$ at middle frequencies, and $\mathrm{St}^{0.4}$ at low frequencies. To the best of the author's knowledge, the PSD of the fluctuating pressure on SUBOFF has not been measured experimentally as a benchmark. Consequently, the present results are consistent with the law from experimental investigations on flows past airfoils and flat plates \citep{Rozenberg2012wallPressure, Hu2016wallPressure, Lee2018WallPressure, Boukharfane2020WallPressure}. Numerical results show that the broadband frequency component is relatively well captured by the WMLES with the high-order numerical model.

\subsubsection{Far-field acoustic signature}\label{suboff_acoustics}
\begin{figure}[!t]
	\centering
	\begin{minipage}[c]{1\linewidth}
		\centering
		\begin{overpic}[width=15cm]{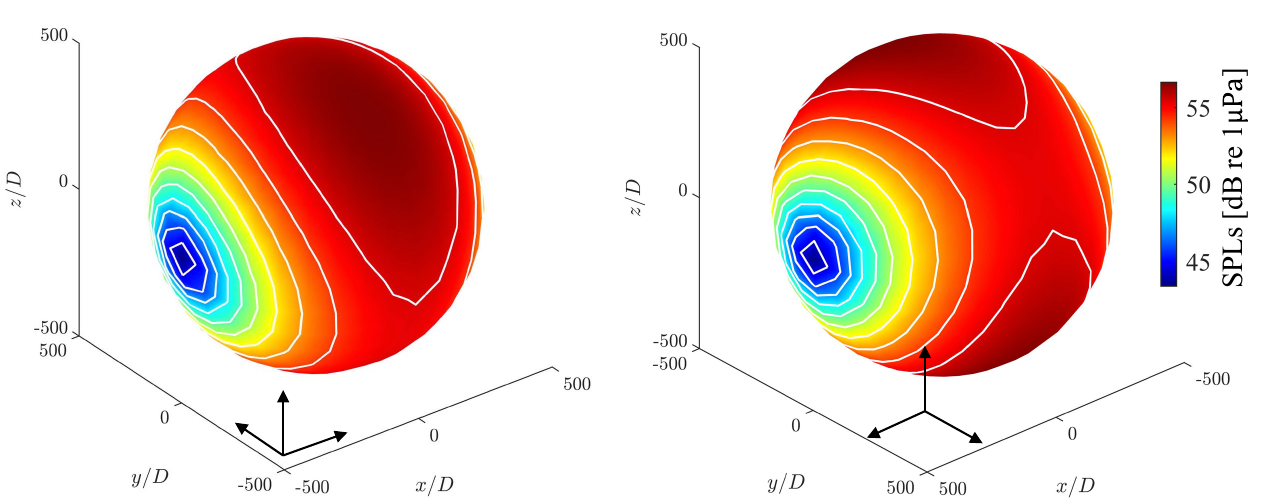}
        \put(30,8){\rotatebox{20}{\includegraphics[scale=0.08]{ Figure_15-1.pdf}}}
        \put(80,15){\rotatebox{-157}{\includegraphics[scale=0.08]{Figure_15-1.pdf}}}
		\put(-1,37){\color{black}{\textit{(a)}}}
        \put(49,36.5){\color{black}{\textit{(b)}}}
        \put(28,6){\color{black}{$x$}}
        \put(17,7){\color{black}{$y$}}
        \put(22,9.5){\color{black}{$z$}}
        \put(78,4){\color{black}{$y$}}
        \put(67,5){\color{black}{$x$}}
        \put(72.5,12.5){\color{black}{$z$}}
		\end{overpic}
	\end{minipage}
	\caption[]{Contours and isolines of the \textit{SPLs} of the root-mean-square sound pressure on a sphere of radial extent 500D, centered on the bow of the SUBOFF hull. View from (a) upstream and (b) downstream. Reference to the orientation of the SUBOFF hull is given in the insets.}
	\label{fig:suboff_acoustics_directivity}
\end{figure}

The distribution of \textit{SPLs} in the far field is acquired from the sphere with 648 hydrophones. As shown in Fig.~\ref{fig:suboff_acoustics_directivity}, both upstream and downstream views are depicted with the orientation of the SUBOFF hull. The following observations can be made: (i) The lowest acoustic pressure levels of 43.37 dB occur in the upstream and downstream directions, while the highest levels of 56.66 dB are observed on the middle-parallel plane. (ii) Similar levels of acoustic pressure characterize the port and starboard sides. (iii) Comparable levels of acoustic pressure are observed in the upstream and downstream directions.

\begin{figure}[!t]
	\centering
 	\begin{minipage}[c]{0.4\linewidth}
		\centering
		\begin{overpic}[width=6cm]{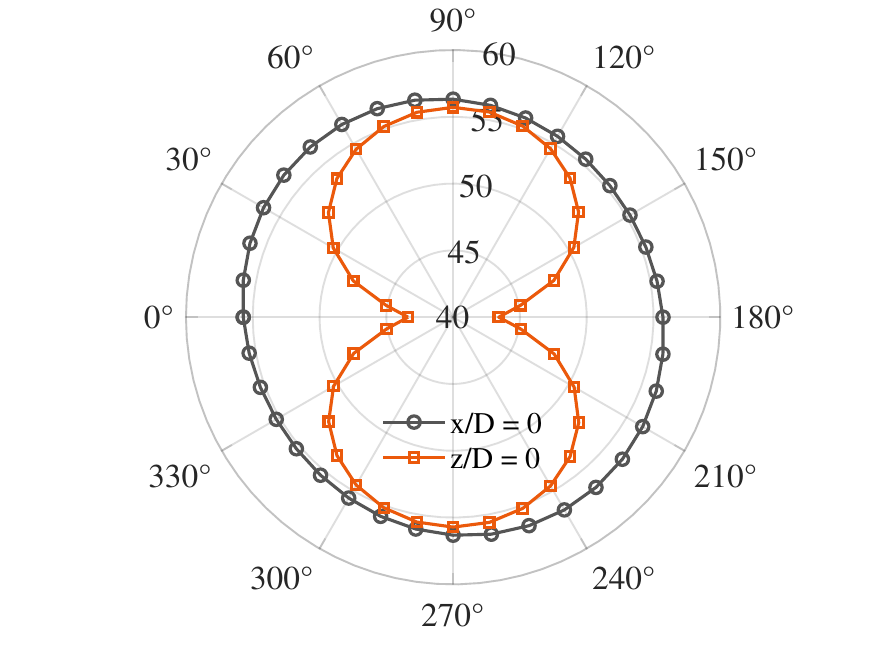}
			\put(0,60){\color{black}{\textit{(a)}}}
            \put(59,68){\rotatebox{-5}{\scriptsize\color{black}{[dB]}}}
            \put(49,65){\footnotesize\color{red}{$P4$}}
            \put(50.3,60){\footnotesize\color{red}{*}}
            \put(63,38){\footnotesize\color{red}{$P5$}}
            \put(55.5,36){\footnotesize\color{red}{*}}
		\end{overpic}	
	\end{minipage}			
	\begin{minipage}[c]{0.55\linewidth}	
		\begin{overpic}[width=9cm]{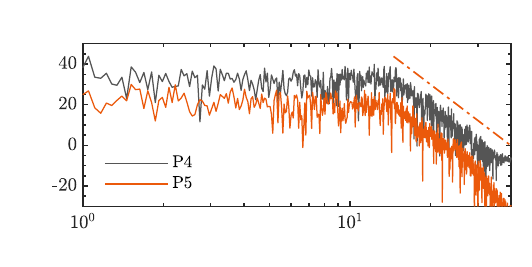}
			\put(-3,48){\color{black}{\textit{(b)}}}
            \put(83,35){\footnotesize{\color{red}{\rotatebox{-30}{$\propto \mathrm{St}^{-5}$}}}}
            \put(2,5){\footnotesize{\color{black}{\rotatebox{90}{$10\log _{10}(\phi_{p^\prime p^\prime}/p^2_{ref})\ \mathrm{[dB/Hz]}$}}}}
            \put(48,2){\footnotesize{\color{black}{St=$fD_c/U_\infty$}}}
		\end{overpic}	
	\end{minipage}
	\caption[]{(a) Far-field noise directivities calculated from the root-mean-square sound pressure at slice $x/D=0$ and $z/D=0$; and (b) the PSD of the far-field sound pressure fluctuations at different locations over $\mathrm{P4}\ (x/D,y/D,z/D) = (0,500,0)$ and $\mathrm{P5}\ (x/D,y/D,z/D) = (500,0,0)$. Refer to Fig.~\ref{fig:suboff_directivity} (a) for the location schematic of probes, which are the points with the highest and lowest sound pressure in the far field at slice $z/D=0$.}
	\label{fig:suboff_directivity}
\end{figure}
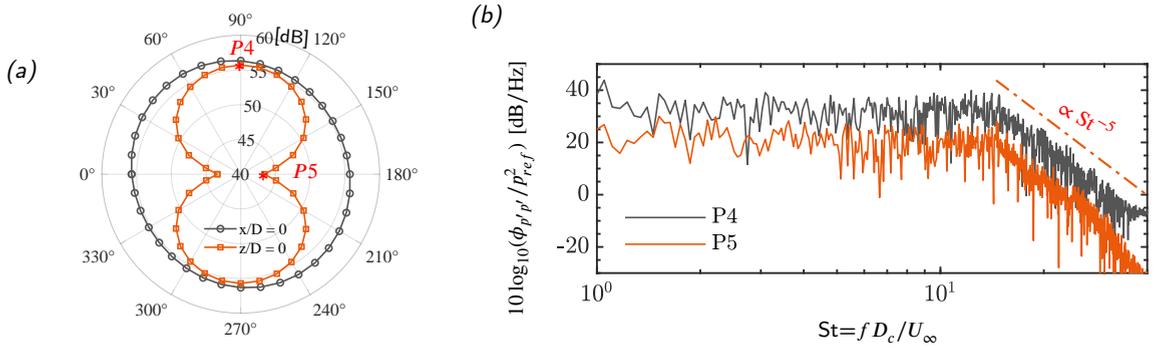

We make further comparisons of far-field noise directivities in polar coordinates which are plotted in 
Fig.~\ref{fig:suboff_directivity} on the planes $x/D = 0$ and $z/D = 0$. The PSD results for the locations of minimum and maximum acoustic pressure are also included for reference. The key findings are as follows: (i) \textit{SPLs} exhibit remarkable consistency in the region of high acoustic pressure at the transverse plane of $x/D=0$ in the hull center with a maximum variation of $1.2$ dB between locations at different angles, which is reasonable for an axisymmetric body of revolution; (ii) In the vertical plane at $z/D=0$, the directivity plot reveals a symmetric dipole pattern about the streamwise and vertical directions, with vertical fluctuations associated with lift stronger than those in the streamwise direction related to drag; (iii) The spectrum of surface fluctuating pressure varies approximately as $\mathrm{St}^{-5}$ at high frequencies; (iv) Both hydrophones exhibit a low-frequency broadband spectral, with the majority of the energy below $\mathrm{St} = 15$.

\section{Conclusions}\label{conclusions}

In this paper, we develop a hybrid numerical model that combines 
the WMLES and the FW--H acoustic analogy based on the high-order FVMS3 scheme.
The present numerical framework is first verified by simulating the flow past a circular cylinder at $\mathrm{Re}=1.0\times10^4$. Our numerical results are carefully examined for both flow statistics and far-field acoustic signatures which show closer agreement with experimental data than conventional second-order schemes. The validated results indicate the hybrid method of high-order CFD and acoustic analogy accurately reproduces the principal flow characteristics and sound propagation mechanisms, establishing the groundwork for the upcoming investigation of turbulent noise of flow over an axisymmetric hull.

After validation, we investigate the hydrodynamic noise of turbulent flows over the DARPA SUBOFF hull geometry by the present model at Reynolds number of $\mathrm{Re} = 1.2\times10^7$. The simulation is carried out on the unstructured grid of 89 million cells which is more extensive than previous WMLES of the SUBOFF at the same $\mathrm{Re}$. \textcolor{black}{The time-averaged surface pressure coefficients, surface skin friction coefficients, mean drag coefficient, and first- and second-order statistics of velocity, predicted by our simulation show good agreement with the existing experimental and numerical results. While traditional schemes struggle to calculate the wall shear stress and drag force of the SUBOFF hull, the present model with the high-order scheme accurately reproduces these turbulent flow characteristics around an axisymmetric hull at $\mathrm{Re}=1.2\times10^7$ which is superior to previous results.
This demonstrates that the present high-order schemes with WMLES provide significant advantages for solving complex high-Reynolds-number turbulence problems.}

We further examined the PSD of wall pressure fluctuations which are a significant source of far-field noise. The PSD reveals a low-frequency broadband spectrum, with the majority of signal energy below $\mathrm{St}= 10$. As evidenced by the slopes displayed on the PSD, the surface fluctuating pressure spectrum follows an approximate variation of \(\mathrm{St}^{-9}\) at high frequencies; the surface fluctuating pressure spectrum varies approximately as $\mathrm{St}^{-3}$ at intermediate frequencies; and the surface fluctuating pressure spectrum varies approximately as $\mathrm{St}^{0. 4}$ at low frequencies. The scaling law is consistent with the scaling of previous studies of flows over airfoils, flat plates, and the SUBOFF model \citep{Rozenberg2012wallPressure, Hu2016wallPressure, Lee2018WallPressure, Boukharfane2020WallPressure, Zhou2022suboff}. In summary, all spectrums show that the broadband frequency component is relatively well captured by the WMLES model with the high-order schemes.

\textcolor{black}{The FW--H acoustic analogy was used to predict the sound distribution in the far field of flow over the bare hull of SUBOFF, at a distance from it equal to 500 hull diameters. Below is a summary of the main characteristics of the far-field acoustic pressure distribution in all spatial directions for the SUBOFF hull:
\begin{enumerate}[(i).]
\item  The lowest sound pressure levels (\textit{SPLs}) of 43.37 dB occur in the upstream and downstream directions while the highest levels of 56.66 dB are found in the mid-parallel plane. Similar levels of sound pressure are present on both the port and starboard sides, and also the upstream and downstream directions.
\item The \textit{SPLs} are extremely close in the region of high acoustic pressure with a maximum difference of $1.2$ dB between locations at different angles, which is reasonable for such an axisymmetric body of revolution. 
\item In the vertical plane at $z/D=0$, the directivity plot reveals a symmetric dipole pattern with vertical fluctuations associated with lift stronger than those in the streamwise direction related to drag.
\item The spectrum of far-field acoustic pressure varies approximately as $\mathrm{St}^{-5}$ at high frequencies, and the PSD of both hydrophones exhibit a low-frequency broadband spectral, with the majority of the energy below $\mathrm{St} = 15$.
\end{enumerate}}

\printcredits

\section*{Declaration of competing interest}
The authors declare that they have no known competing financial
interests or personal relationships that could have appeared to influence the work reported in this paper.

\section*{Data availability}
The data that support the findings of this study are available from the corresponding author upon reasonable request.

\section*{Acknowledgements}

We would like to thank Dr G.J. Chen for sharing the data of the flow past a cylinder.
This work was supported in part by Shanghai Rising-Star Program (No. 23QA1405000), Shanghai Pilot Program for Basic Research - Shanghai Jiao Tong University (No. 21TQ1400202), the fund from the Fundamental Research Funds for the Central Universities as well as the support from National Natural Science Foundation of China (No. 91752104).

\bibliographystyle{cas-model2-names}


\begin{thebibliography}{57}
\expandafter\ifx\csname natexlab\endcsname\relax\def\natexlab#1{#1}\fi
\providecommand{\url}[1]{\texttt{#1}}
\providecommand{\href}[2]{#2}
\providecommand{\path}[1]{#1}
\providecommand{\DOIprefix}{doi:}
\providecommand{\ArXivprefix}{arXiv:}
\providecommand{\URLprefix}{URL: }
\providecommand{\Pubmedprefix}{pmid:}
\providecommand{\doi}[1]{\href{http://dx.doi.org/#1}{\path{#1}}}
\providecommand{\Pubmed}[1]{\href{pmid:#1}{\path{#1}}}
\providecommand{\bibinfo}[2]{#2}
\ifx\xfnm\relax \def\xfnm[#1]{\unskip,\space#1}\fi
\bibitem[{Boukharfane et~al.(2020)Boukharfane, Parsani and Bodart}]{Boukharfane2020WallPressure}
\bibinfo{author}{Boukharfane, R.}, \bibinfo{author}{Parsani, M.}, \bibinfo{author}{Bodart, J.}, \bibinfo{year}{2020}.
\newblock \bibinfo{title}{Characterization of pressure fluctuations within a controlled-diffusion blade boundary layer using the equilibrium wall-modelled {LES}}.
\newblock \bibinfo{journal}{Sci.~Rep.} \bibinfo{volume}{10}.
\bibitem[{Brentner and Farassat(2003)}]{Brentner2003rotors}
\bibinfo{author}{Brentner, K.}, \bibinfo{author}{Farassat, F.}, \bibinfo{year}{2003}.
\newblock \bibinfo{title}{Modeling aerodynamically generated sound of helicopter rotors}.
\newblock \bibinfo{journal}{Prog.~Aerospace Sci.} \bibinfo{volume}{39}, \bibinfo{pages}{83--120}.
\bibitem[{Chase and Carrica(2013)}]{Chase2013OE}
\bibinfo{author}{Chase, N.}, \bibinfo{author}{Carrica, P.M.}, \bibinfo{year}{2013}.
\newblock \bibinfo{title}{Submarine propeller computations and application to self-propulsion of {DARPA Suboff}}.
\newblock \bibinfo{journal}{Ocean Eng.} \bibinfo{volume}{60}, \bibinfo{pages}{68--80}.
\bibitem[{Chen et~al.(2023a)Chen, Zang and Azarpeyvand}]{Chen2023Cylinder}
\bibinfo{author}{Chen, G.J.}, \bibinfo{author}{Zang, B.}, \bibinfo{author}{Azarpeyvand, M.}, \bibinfo{year}{2023}a.
\newblock \bibinfo{title}{Numerical investigation on aerodynamic noise of flow past a cylinder with different spanwise lengths}.
\newblock \bibinfo{journal}{Phys.~Fluids} \bibinfo{volume}{35}, \bibinfo{pages}{035128}.
\bibitem[{Chen et~al.(2023b)Chen, Yang, Zhao and Wan}]{Chen2023SUBOFF}
\bibinfo{author}{Chen, S.T.}, \bibinfo{author}{Yang, L.C.}, \bibinfo{author}{Zhao, W.W.}, \bibinfo{author}{Wan, D.C.}, \bibinfo{year}{2023}b.
\newblock \bibinfo{title}{Wall-modeled large eddy simulation for the flows around an axisymmetric body of revolution}.
\newblock \bibinfo{journal}{J.~Hydrodyn.} \bibinfo{volume}{35}, \bibinfo{pages}{199--209}.
\bibitem[{Choi and Moin(2012)}]{Choi2012gridWMLES}
\bibinfo{author}{Choi, H.}, \bibinfo{author}{Moin, P.}, \bibinfo{year}{2012}.
\newblock \bibinfo{title}{Grid-point requirements for large eddy simulation: {Chapman's} estimates revisited}.
\newblock \bibinfo{journal}{Phys.~Fluids} \bibinfo{volume}{24}, \bibinfo{pages}{011702}.
\bibitem[{Chorin(1968)}]{Chorin1968Numerical}
\bibinfo{author}{Chorin, A.J.}, \bibinfo{year}{1968}.
\newblock \bibinfo{title}{Numerical solution of the {Navier-Stokes} equations}.
\newblock \bibinfo{journal}{Math.~Comput.} \bibinfo{volume}{22}, \bibinfo{pages}{745--762}.
\bibitem[{Cianferra et~al.(2019)Cianferra, Petronio and Armenio}]{Cianferra2019shipNoise}
\bibinfo{author}{Cianferra, M.}, \bibinfo{author}{Petronio, A.}, \bibinfo{author}{Armenio, V.}, \bibinfo{year}{2019}.
\newblock \bibinfo{title}{Non-linear noise from a ship propeller in open sea conditions}.
\newblock \bibinfo{journal}{Ocean Eng.} \bibinfo{volume}{191}, \bibinfo{pages}{106474}.
\bibitem[{Curle(1955)}]{Curle1955}
\bibinfo{author}{Curle, N.}, \bibinfo{year}{1955}.
\newblock \bibinfo{title}{The influence of solid boundaries upon aerodynamic sound}.
\newblock \bibinfo{journal}{Proc.~R.~Soc.~Lond.~A} \bibinfo{volume}{231}, \bibinfo{pages}{505--514}.
\bibitem[{Dong and Karniadakis(2005)}]{Dong2005DNS}
\bibinfo{author}{Dong, S.}, \bibinfo{author}{Karniadakis, G.E.}, \bibinfo{year}{2005}.
\newblock \bibinfo{title}{{DNS} of flow past a stationary and oscillating cylinder at {$Re$}=10000}.
\newblock \bibinfo{journal}{J.~Fluids Struct.} \bibinfo{volume}{20}, \bibinfo{pages}{519–531}.
\bibitem[{Duprat et~al.(2011)Duprat, Balarac, Métais, Congedo and Brugière}]{Duprat2011nutmodel}
\bibinfo{author}{Duprat, C.}, \bibinfo{author}{Balarac, G.}, \bibinfo{author}{Métais, O.}, \bibinfo{author}{Congedo, P.}, \bibinfo{author}{Brugière, O.}, \bibinfo{year}{2011}.
\newblock \bibinfo{title}{A wall-layer model for large-eddy simulations of turbulent flows with/out pressure gradient}.
\newblock \bibinfo{journal}{Phys.~Fluids} \bibinfo{volume}{23}, \bibinfo{pages}{015101}.
\bibitem[{Farassat(2007)}]{Farassat20071A}
\bibinfo{author}{Farassat, F.}, \bibinfo{year}{2007}.
\newblock \bibinfo{title}{Derivation of formulations 1 and {1A} of {Farassat}}.
\newblock \bibinfo{type}{Technical Report} \bibinfo{number}{NASA/TM-2007-214853}. Langley Research Center. \bibinfo{address}{Hampton, Virginia}.
\bibitem[{{Ffowcs Williams} and Hawkings(1969)}]{FWH1969}
\bibinfo{author}{{Ffowcs Williams}, J.E.}, \bibinfo{author}{Hawkings, D.L.}, \bibinfo{year}{1969}.
\newblock \bibinfo{title}{Sound generation by turbulence and surfaces in arbitrary motion}.
\newblock \bibinfo{journal}{Philos.~Trans.~R.~Soc.~London~Ser.~A} \bibinfo{volume}{264}, \bibinfo{pages}{321--342}.
\bibitem[{Gopalkrishnan(1993)}]{Gopalkrishnan1993DNS}
\bibinfo{author}{Gopalkrishnan, R.}, \bibinfo{year}{1993}.
\newblock \bibinfo{title}{Vortex-induced forces on oscillating bluff cylinders}.
\newblock \bibinfo{type}{Ph.{D}.}. Massachusetts Institute of Technology.
\bibitem[{Gottlieb and Shu(1998)}]{Gottlieb1998TVD}
\bibinfo{author}{Gottlieb, S.}, \bibinfo{author}{Shu, C.W.}, \bibinfo{year}{1998}.
\newblock \bibinfo{title}{Total variation diminishing {Runge-Kutta} schemes}.
\newblock \bibinfo{journal}{Math.~Comput.} \bibinfo{volume}{67}, \bibinfo{pages}{73--85}.
\bibitem[{Groves et~al.(1989)Groves, Huang and Chang}]{groves1989geometric}
\bibinfo{author}{Groves, N.C.}, \bibinfo{author}{Huang, T.T.}, \bibinfo{author}{Chang, M.S.}, \bibinfo{year}{1989}.
\newblock \bibinfo{title}{{Geometric characteristics of DARPA (Defense Advanced Research Projects Agency) SUBOFF models (DTRC model numbers 5470 and 5471)}}.
\newblock \bibinfo{type}{SHD-1298-01}. David Taylor Research Center.
\bibitem[{Hayat and Park(2023)}]{Hayat2023WMLES}
\bibinfo{author}{Hayat, I.}, \bibinfo{author}{Park, G.I.}, \bibinfo{year}{2023}.
\newblock \bibinfo{title}{Efficient spectral implementation of ode wall model and the extension of integral wall model to unstructured {LES} solvers}.
\newblock \bibinfo{journal}{J.~Comput.~Phys.} \bibinfo{volume}{487}, \bibinfo{pages}{112175}.
\bibitem[{He et~al.(2024)He, Pan, Zhao, Wang and Wan}]{He2024Review}
\bibinfo{author}{He, K.J.}, \bibinfo{author}{Pan, Z.}, \bibinfo{author}{Zhao, W.W.}, \bibinfo{author}{Wang, J.H.}, \bibinfo{author}{Wan, D.C.}, \bibinfo{year}{2024}.
\newblock \bibinfo{title}{Overview of research progress on numerical simulation methods for turbulent flows around underwater vehicles}.
\newblock \bibinfo{journal}{J.~Marine.~Sci.~Appl.} \bibinfo{volume}{23}, \bibinfo{pages}{1--22}.
\bibitem[{He et~al.(2023)He, Zhou, Zhao, Wang and Wan}]{He2023SUBOFF}
\bibinfo{author}{He, K.J.}, \bibinfo{author}{Zhou, F.C.}, \bibinfo{author}{Zhao, W.W.}, \bibinfo{author}{Wang, J.H.}, \bibinfo{author}{Wan, D.C.}, \bibinfo{year}{2023}.
\newblock \bibinfo{title}{Numerical analysis of turbulent fluctuations around an axisymmetric body of revolution based on wall-modeled large eddy simulations}.
\newblock \bibinfo{journal}{J.~Hydrodyn.} \bibinfo{volume}{35}, \bibinfo{pages}{1041–1051}.
\bibitem[{Hu and Herr(2016)}]{Hu2016wallPressure}
\bibinfo{author}{Hu, N.}, \bibinfo{author}{Herr, M.}, \bibinfo{year}{2016}.
\newblock \bibinfo{title}{Characteristics of wall pressure fluctuations for a flat plate turbulent boundary layer with pressure gradients}, in: \bibinfo{booktitle}{AIAA/CEAS 22nd Aeroacoustics Conference, AIAA Paper}, p. \bibinfo{pages}{2749}.
\bibitem[{Hu et~al.(2023)Hu, Hayat and Park}]{Hu2023WMLES}
\bibinfo{author}{Hu, X.}, \bibinfo{author}{Hayat, I.}, \bibinfo{author}{Park, G.I.}, \bibinfo{year}{2023}.
\newblock \bibinfo{title}{Wall-modelled large-eddy simulation of three-dimensional turbulent boundary layer in a bent square duct}.
\newblock \bibinfo{journal}{J.~Fluid Mech.} \bibinfo{volume}{960}, \bibinfo{pages}{A29}.
\bibitem[{Huang et~al.(1992)Huang, Liu, Groves, Forlini, Blanton and Gowing}]{Huang1992Exp}
\bibinfo{author}{Huang, T.}, \bibinfo{author}{Liu, H.L.}, \bibinfo{author}{Groves, N.C.}, \bibinfo{author}{Forlini, T.}, \bibinfo{author}{Blanton, J.}, \bibinfo{author}{Gowing, S.}, \bibinfo{year}{1992}.
\newblock \bibinfo{title}{Measurements of flows over an axisymmetric body with various appendages in a wind tunnel: {The DARPA SUBOFF} experimental program}, in: \bibinfo{booktitle}{In Proceedings of the 19th Symposium on Naval Hydrodynamics}.
\bibitem[{Hunt et~al.(1988)Hunt, Wray and Moin}]{hunt1988eddies}
\bibinfo{author}{Hunt, J.C.R.}, \bibinfo{author}{Wray, A.A.}, \bibinfo{author}{Moin, P.}, \bibinfo{year}{1988}.
\newblock \bibinfo{title}{Eddies, streams, and convergence zones in turbulent flows}.
\newblock \bibinfo{type}{CTR S88}. Center for Turbulence Research Report. \bibinfo{address}{Stanford}.
\bibitem[{Jiménez et~al.(2010)Jiménez, Hoyas, Simens and Mizuno}]{TBL2010}
\bibinfo{author}{Jiménez, J.}, \bibinfo{author}{Hoyas, S.}, \bibinfo{author}{Simens, M.P.}, \bibinfo{author}{Mizuno, Y.}, \bibinfo{year}{2010}.
\newblock \bibinfo{title}{Turbulent boundary layers and channels at moderate {Reynolds} numbers}.
\newblock \bibinfo{journal}{J.~Fluid Mech.} \bibinfo{volume}{657}, \bibinfo{pages}{335--360}.
\bibitem[{Kato et~al.(1993)Kato, Iida, Takano, Fujita and Ikegawa}]{kato1993numerical}
\bibinfo{author}{Kato, C.}, \bibinfo{author}{Iida, A.}, \bibinfo{author}{Takano, Y.}, \bibinfo{author}{Fujita, H.}, \bibinfo{author}{Ikegawa, M.}, \bibinfo{year}{1993}.
\newblock \bibinfo{title}{Numerical prediction of aerodynamic noise radiated from low {Mach} number turbulent wake}, in: \bibinfo{booktitle}{31st aerospace sciences meeting}, p. \bibinfo{pages}{145}.
\bibitem[{Kawai and Larsson(2012)}]{Kawai2012WallmodelingIL}
\bibinfo{author}{Kawai, S.}, \bibinfo{author}{Larsson, J.}, \bibinfo{year}{2012}.
\newblock \bibinfo{title}{Wall-modeling in large eddy simulation: Length scales, grid resolution, and accuracy}.
\newblock \bibinfo{journal}{Phys.~Fluids} \bibinfo{volume}{24}, \bibinfo{pages}{015105}.
\bibitem[{Kim and Moin(1985)}]{Kim1985Application}
\bibinfo{author}{Kim, J.}, \bibinfo{author}{Moin, P.}, \bibinfo{year}{1985}.
\newblock \bibinfo{title}{Application of a fractional-step method to incompressible {Navier-Stokes} equations}.
\newblock \bibinfo{journal}{J.~Comput.~Phys.} \bibinfo{volume}{59}, \bibinfo{pages}{308--323}.
\bibitem[{Kumar and Mahesh(2018)}]{Kumar2018SUBOFF}
\bibinfo{author}{Kumar, P.}, \bibinfo{author}{Mahesh, K.}, \bibinfo{year}{2018}.
\newblock \bibinfo{title}{Large-eddy simulation of flow over an axisymmetric body of revolution}.
\newblock \bibinfo{journal}{J.~Fluid Mech.} \bibinfo{volume}{853}, \bibinfo{pages}{537--563}.
\bibitem[{Le and Moin(1991)}]{LE1991369}
\bibinfo{author}{Le, H.}, \bibinfo{author}{Moin, P.}, \bibinfo{year}{1991}.
\newblock \bibinfo{title}{An improvement of fractional step methods for the incompressible navier-stokes equations}.
\newblock \bibinfo{journal}{J.~Comput.~Phys.} \bibinfo{volume}{92}, \bibinfo{pages}{369--379}.
\bibitem[{Lee and Moser(2015)}]{Lee2015Channel}
\bibinfo{author}{Lee, M.}, \bibinfo{author}{Moser, R.D.}, \bibinfo{year}{2015}.
\newblock \bibinfo{title}{Direct numerical simulation of turbulent channel flow up to {$Re_\tau\approx$}5200}.
\newblock \bibinfo{journal}{J.~Fluid Mech.} \bibinfo{volume}{774}, \bibinfo{pages}{395--415}.
\bibitem[{Lee(2018)}]{Lee2018WallPressure}
\bibinfo{author}{Lee, S.}, \bibinfo{year}{2018}.
\newblock \bibinfo{title}{Empirical wall-pressure spectral modeling for zero and adverse pressure gradient flows}.
\newblock \bibinfo{journal}{{AIAA} J.} \bibinfo{volume}{56}.
\bibitem[{Li et~al.(2016)Li, Liu, Wu and Chen}]{Li2016review}
\bibinfo{author}{Li, H.}, \bibinfo{author}{Liu, C.W.}, \bibinfo{author}{Wu, F.l.}, \bibinfo{author}{Chen, C.}, \bibinfo{year}{2016}.
\newblock \bibinfo{title}{A review of the progress for computational methods of hydrodynamic noise}.
\newblock \bibinfo{journal}{J.~Ship Res.} \bibinfo{volume}{11}, \bibinfo{pages}{72--89}.
\newblock \bibinfo{note}{(in Chinese)}.
\bibitem[{Lighthill(1952)}]{Lighthill1952I}
\bibinfo{author}{Lighthill, M.}, \bibinfo{year}{1952}.
\newblock \bibinfo{title}{On sound generated aerodynamically\text{-}{I}: {General} theory}.
\newblock \bibinfo{journal}{Proc.~R.~Soc.~Lond.~A} \bibinfo{volume}{211}, \bibinfo{pages}{564–587}.
\bibitem[{Lighthill(1954)}]{Lighthill1954II}
\bibinfo{author}{Lighthill, M.}, \bibinfo{year}{1954}.
\newblock \bibinfo{title}{On sound generated aerodynamically\text{-}{II}: {Turbulence} as a source of sound}.
\newblock \bibinfo{journal}{Proc.~R.~Soc.~Lond.~A} \bibinfo{volume}{222}, \bibinfo{pages}{1--32}.
\bibitem[{Liu et~al.(2023)Liu, Wang, Wang and He}]{LIU2023112009}
\bibinfo{author}{Liu, Y.}, \bibinfo{author}{Wang, H.P.}, \bibinfo{author}{Wang, S.Z.}, \bibinfo{author}{He, G.W.}, \bibinfo{year}{2023}.
\newblock \bibinfo{title}{A cache-efficient reordering method for unstructured meshes with applications to wall-resolved large-eddy simulations}.
\newblock \bibinfo{journal}{J.~Comput.~Phys.} \bibinfo{volume}{480}, \bibinfo{pages}{112009}.
\bibitem[{Morse and Mahesh(2021)}]{morse2021suboff}
\bibinfo{author}{Morse, N.}, \bibinfo{author}{Mahesh, K.}, \bibinfo{year}{2021}.
\newblock \bibinfo{title}{Large-eddy simulation and streamline coordinate analysis of flow over an axisymmetric hull}.
\newblock \bibinfo{journal}{J.~Fluid~Mech.} \bibinfo{volume}{926}, \bibinfo{pages}{A18}.
\bibitem[{Mukha et~al.(2019)Mukha, Rezaeiravesh and Liefvendahl}]{Mukha2019WMLES}
\bibinfo{author}{Mukha, T.}, \bibinfo{author}{Rezaeiravesh, S.}, \bibinfo{author}{Liefvendahl, M.}, \bibinfo{year}{2019}.
\newblock \bibinfo{title}{A library for wall-modelled large-eddy simulation based on {OpenFOAM} technology}.
\newblock \bibinfo{journal}{Comput.~Phys.~Commun.} \bibinfo{volume}{239}, \bibinfo{pages}{204--224}.
\bibitem[{Nicoud and Ducros(1999)}]{Nicoud1999WALE}
\bibinfo{author}{Nicoud, F.}, \bibinfo{author}{Ducros, F.}, \bibinfo{year}{1999}.
\newblock \bibinfo{title}{Subgrid-scale stress modelling based on the square of the velocity gradient tensor}.
\newblock \bibinfo{journal}{Flow~Turbul.~Combust.} \bibinfo{volume}{62}, \bibinfo{pages}{183--200}.
\bibitem[{{Ortiz-Tarin} et~al.(2021){Ortiz-Tarin}, Nidhan,  and Sarkar}]{Ortiz2021slenderBody}
\bibinfo{author}{{Ortiz-Tarin}, J.}, \bibinfo{author}{Nidhan, S.}, , \bibinfo{author}{Sarkar, S.}, \bibinfo{year}{2021}.
\newblock \bibinfo{title}{{High-Reynolds-number} wake of a slender body}.
\newblock \bibinfo{journal}{J.~Fluid~Mech.} \bibinfo{volume}{918}, \bibinfo{pages}{A30}.
\bibitem[{Park and Moin(2016)}]{Park2016wallPressure}
\bibinfo{author}{Park, G.I.}, \bibinfo{author}{Moin, P.}, \bibinfo{year}{2016}.
\newblock \bibinfo{title}{{Space-time characteristics of wall-pressure and wall shear-stress fluctuations in wall-modeled large eddy simulation}}.
\newblock \bibinfo{journal}{Phys.~Rev.~Fluid} \bibinfo{volume}{1}, \bibinfo{pages}{024404}.
\bibitem[{Phillips(1956)}]{Phillips1956}
\bibinfo{author}{Phillips, O.}, \bibinfo{year}{1956}.
\newblock \bibinfo{title}{{The intnesity of Aeolian tones}}.
\newblock \bibinfo{journal}{J.~Fluid~Mech.} \bibinfo{volume}{1}, \bibinfo{pages}{607--624}.
\bibitem[{Posa and Balaras(2016)}]{posa2016suboff}
\bibinfo{author}{Posa, A.}, \bibinfo{author}{Balaras, E.}, \bibinfo{year}{2016}.
\newblock \bibinfo{title}{A numerical investigation of the wake of an axisymmetric body with appendages}.
\newblock \bibinfo{journal}{J.~Fluid~Mech.} \bibinfo{volume}{792}, \bibinfo{pages}{470--498}.
\bibitem[{Posa and Balaras(2020)}]{posa2020suboff}
\bibinfo{author}{Posa, A.}, \bibinfo{author}{Balaras, E.}, \bibinfo{year}{2020}.
\newblock \bibinfo{title}{A numerical investigation about the effects of {Reynolds} number on the flow around an appended axisymmetric body of revolution}.
\newblock \bibinfo{journal}{J.~Fluid~Mech.} \bibinfo{volume}{884}, \bibinfo{pages}{A41}.
\bibitem[{Posa et~al.(2022)Posa, Felli and Broglia}]{Posa2022propeller}
\bibinfo{author}{Posa, A.}, \bibinfo{author}{Felli, M.}, \bibinfo{author}{Broglia, R.}, \bibinfo{year}{2022}.
\newblock \bibinfo{title}{{Influence of an upstream hydrofoil on the acoustic signature of a propeller}}.
\newblock \bibinfo{journal}{Phys.~Fluids} \bibinfo{volume}{34}, \bibinfo{pages}{045112}.
\bibitem[{Posa et~al.(2023)Posa, Felli and Broglia}]{Posa2023propeller}
\bibinfo{author}{Posa, A.}, \bibinfo{author}{Felli, M.}, \bibinfo{author}{Broglia, R.}, \bibinfo{year}{2023}.
\newblock \bibinfo{title}{{Acoustic far field of a propeller working in the wake of a hydrofoil}}.
\newblock \bibinfo{journal}{Phys.~Fluids} \bibinfo{volume}{35}, \bibinfo{pages}{125121}.
\bibitem[{Qu et~al.(2021)Qu, Wu, Zhao, Huang, Fu and Wang}]{Qu2021suboff}
\bibinfo{author}{Qu, Y.}, \bibinfo{author}{Wu, Q.}, \bibinfo{author}{Zhao, X.}, \bibinfo{author}{Huang, B.}, \bibinfo{author}{Fu, X.}, \bibinfo{author}{Wang, G.}, \bibinfo{year}{2021}.
\newblock \bibinfo{title}{Numerical investigation of flow structures around the {DARPA SUBOFF} model}.
\newblock \bibinfo{journal}{Ocean Eng.} \bibinfo{volume}{239}, \bibinfo{pages}{109866}.
\bibitem[{Robison and Peake(2014)}]{Robison2014propeller}
\bibinfo{author}{Robison, R.A.V.}, \bibinfo{author}{Peake, N.}, \bibinfo{year}{2014}.
\newblock \bibinfo{title}{Noise generation by turbulence-propeller interaction in asymmetric flow}.
\newblock \bibinfo{journal}{J.~Fluid~Mech.} \bibinfo{volume}{758}, \bibinfo{pages}{121--149}.
\bibitem[{Rocca et~al.(2022)Rocca, Cianferra, Broglia and Armenio}]{Rocca2022BB2}
\bibinfo{author}{Rocca, A.}, \bibinfo{author}{Cianferra, M.}, \bibinfo{author}{Broglia, R.}, \bibinfo{author}{Armenio, V.}, \bibinfo{year}{2022}.
\newblock \bibinfo{title}{Computational hydroacoustic analysis of the {BB2} submarine using the advective {Ffowcs Williams and Hawkings} equation with {Wall-Modeled LES}}.
\newblock \bibinfo{journal}{Applied Ocean Research} \bibinfo{volume}{129}, \bibinfo{pages}{103360}.
\bibitem[{Rozenberg et~al.(2012)Rozenberg, Robert and Moreau}]{Rozenberg2012wallPressure}
\bibinfo{author}{Rozenberg, Y.}, \bibinfo{author}{Robert, G.}, \bibinfo{author}{Moreau, S.}, \bibinfo{year}{2012}.
\newblock \bibinfo{title}{Wall-pressure spectral model including the adverse pressure gradient effects}.
\newblock \bibinfo{journal}{{AIAA} J.} \bibinfo{volume}{50}, \bibinfo{pages}{2168--2179}.
\bibitem[{Sezen et~al.(2018)Sezen, A. and C.}]{Sezen2018OE}
\bibinfo{author}{Sezen, S.}, \bibinfo{author}{A., D.}, \bibinfo{author}{C., D.}, \bibinfo{year}{2018}.
\newblock \bibinfo{title}{Investigation of self-propulsion of {DARPA Suboff} by {RANS} method}.
\newblock \bibinfo{journal}{Ocean Eng.} \bibinfo{volume}{150}, \bibinfo{pages}{258--271}.
\bibitem[{Wang et~al.(2006)Wang, Freund and Lele}]{Wang2006noiseReview}
\bibinfo{author}{Wang, M.}, \bibinfo{author}{Freund, J.B.}, \bibinfo{author}{Lele, S.K.}, \bibinfo{year}{2006}.
\newblock \bibinfo{title}{Computational prediction of flow-generated sound}.
\newblock \bibinfo{journal}{Annu.~Rev.~Fluid Mech.} \bibinfo{volume}{38}, \bibinfo{pages}{483--512}.
\bibitem[{Welch(1967)}]{Welch1967}
\bibinfo{author}{Welch, P.}, \bibinfo{year}{1967}.
\newblock \bibinfo{title}{The use of fast {Fourier} transform for the estimation of power spectra: a method based on time averaging over short, modified periodograms}.
\newblock \bibinfo{journal}{IEEE trans.~audio electroacoust.} \bibinfo{volume}{15}, \bibinfo{pages}{70--73}.
\bibitem[{Xie et~al.(2019)Xie, Deng and Liao}]{Xie2019High-fidelitysolver}
\bibinfo{author}{Xie, B.}, \bibinfo{author}{Deng, X.}, \bibinfo{author}{Liao, S.J.}, \bibinfo{year}{2019}.
\newblock \bibinfo{title}{High-fidelity solver on polyhedral unstructured grids for low-{Mach} number compressible viscous flow}.
\newblock \bibinfo{journal}{Comput.~Methods Appl.~Mech.~Eng.} \bibinfo{volume}{357}.
\bibitem[{Xie et~al.(2020)Xie, Jin, Du and Liao}]{Xie2020consistent}
\bibinfo{author}{Xie, B.}, \bibinfo{author}{Jin, P.}, \bibinfo{author}{Du, Y.P.}, \bibinfo{author}{Liao, S.J.}, \bibinfo{year}{2020}.
\newblock \bibinfo{title}{A consistent and balanced-force model for incompressible multiphase flows on polyhedral unstructured grids}.
\newblock \bibinfo{journal}{Int.~J.~Multiph.~Flow} \bibinfo{volume}{122}, \bibinfo{pages}{103125}.
\bibitem[{Xie and Xiao(2017)}]{xie2017piso}
\bibinfo{author}{Xie, B.}, \bibinfo{author}{Xiao, F.}, \bibinfo{year}{2017}.
\newblock \bibinfo{title}{Accurate and robust {PISO} algorithm on hybrid unstructured grids using the multi-moment finite volume method}.
\newblock \bibinfo{journal}{Numer.~Heat.~Tr.~B-Fund.} \bibinfo{volume}{71}, \bibinfo{pages}{146--172}.
\bibitem[{Yu et~al.(2007)Yu, Wu and Pang}]{Yu2007shipNoise}
\bibinfo{author}{Yu, M.S.}, \bibinfo{author}{Wu, Y.S.}, \bibinfo{author}{Pang, Y.Z.}, \bibinfo{year}{2007}.
\newblock \bibinfo{title}{A review of progress for hydrodynamic noise of ships}.
\newblock \bibinfo{journal}{J.~Ship Mech.} \bibinfo{volume}{11}, \bibinfo{pages}{152--158}.
\newblock \bibinfo{note}{(in Chinese)}.
\bibitem[{Zhou et~al.(2022)Zhou, Xu, Wang and He}]{Zhou2022suboff}
\bibinfo{author}{Zhou, Z.T.}, \bibinfo{author}{Xu, Z.Y.}, \bibinfo{author}{Wang, S.Z.}, \bibinfo{author}{He, G.W.}, \bibinfo{year}{2022}.
\newblock \bibinfo{title}{{Wall-modeled large-eddy simulation of noise generated by turbulence around an appended axisymmetric body of revolution}}.
\newblock \bibinfo{journal}{J.~Hydrodynam.~B} \bibinfo{volume}{34}, \bibinfo{pages}{533--554}.

\end{thebibliography}

\end{document}